\documentclass[twocolumn,amsmath,amssymb,nofootinbib,eqsecnum,tighten,prb,subentry]{revtex4-1}

\usepackage{graphicx, subcaption}                       
\usepackage[pdftex]{hyperref}                           

\usepackage{amsthm,mathtools,nicefrac,bm,bbm,relsize}   
\newcommand{\A} {\mathcal{A}}
\newcommand{\B} {\mathcal{B}}
\newcommand{\Q} {\mathcal{Q}}
\newcommand{\RR}{\mathcal{R}}
\newcommand{\II}{\mathcal{I}}

\usepackage{xcolor}                                     
\definecolor{mymagenta}{rgb}{0.75, 0, 0.75}
\definecolor{myyellow}{RGB}{255, 214, 00}
\definecolor{myturquois}{rgb}{0, 0.75, 0.75}

\begin{document}


\title{Color-dependent interactions in the three coloring model}
\date{\today}

\author{Philipp C. Verpoort$^{1,2}$}
\author{Jacob Simmons$^{3}$}
\author{Claudio Castelnovo$^{1,4,5}$}

\affiliation{$^1$
Department of Physics, University of Cambridge, J.J. Thomson Avenue,
Cambridge CB3 0HE, UK
}
\affiliation{$^2$
Karlsruhe Institute of Technology, Institute for Theory of Condensed Matter,
D-76131 Karlsruhe, Germany
}
\affiliation{$^3$
Maine Maritime Academy,
Pleasant Street,
Castine, ME 04420, USA
}
\affiliation{$^4$
Rudolf Peierls Centre for Theoretical Physics, University of Oxford,
Oxford, OX1 3NP, United Kingdom
}
\affiliation{$^5$
SEPnet and Hubbard Theory Consortium, Department of Physics,
Royal Holloway University of London,
Egham TW20 0EX, United Kingdom
}


\begin{abstract}
Since it was first discussed by Baxter in 1970, the three coloring model has
been studied in several contexts, from frustrated magnetism to superconducting
devices and glassiness. In presence of interactions, when the model is no
longer exactly soluble, it was already observed that the phase diagram is
highly non-trivial. Here we discuss the generic case of `color-dependent'
nearest-neighbor interactions between the vertex chiralities. We uncover
different critical regimes merging into one another: $c=\nicefrac{1}{2}$ free
fermions combining into $c=1$ free bosons; $c=1$ free bosons combining into
$c=2$ critical loop models; as well as three separate $c=\nicefrac{1}{2}$
critical lines merging at a supersymmetric $c=\nicefrac{3}{2}$
critical point. When the three coupling constants are tuned to equal one
another, transfer-matrix calculations highlight a puzzling regime where the
central charge appears to vary continuously from $\nicefrac{3}{2}$ to $2$.
\end{abstract}

\maketitle


\section{Introduction}
\label{sec: intro}

The three coloring model was introduced by Baxter in 1970 as the combinatorial
problem to compute the ``number of ways ... of coloring the bonds of a
hexagonal lattice ... with three colors so that no adjacent bonds are colored
alike''. The author showed that the model is integrable in the absence of
interactions and proceeded to find an exact solution~\cite{Baxter1970}.

In the absence of interactions, a parallel can be drawn between the three
coloring model and the fully packed loop model with fugacity $2$. The latter is
critical and was argued to have central charge $c=2$ and SU(3) symmetry by
Reshetikhin~\cite{Reshetikhin1991}, from results on the integrability of the
model. At a kagome workshop in 1992, N.~Read presented a formulation of the
model that illustrates explicitly the SU(3) symmetry and argued that the
long-wavelength limit is described by an SU(3)$_1$ conformal field
theory~\footnote{The very elegant argument put forward by N.~Read was never
published and we report it for completeness in App.~\ref{app: Read}, with his
kind permission.}, by means of mapping to a two-component
height model~\cite{Huse1992}. This was later confirmed by Kondev and
collaborators~\cite{Kondev1996-1,*Kondev1996-2}.

The model received renewed attention when it was noticed that three coloring
configurations describe the ground states of an Heisenberg antiferromagnet on
the kagome lattice~\cite{Huse1992}.
The physics of the three coloring model was also found to be relevant to
the behavior of arrays of Josephson junctions~\cite{Moore2003,Castelnovo2004}
and kagome networks of superconducting
wires~\cite{Higgins2000,Park2001,Xiao2002,Castelnovo2004}, provided that
appropriate (uniform) interactions are introduced in the model. These
interactions -- which are typically written in terms of vertex chirality spins,
encoding the parity of the three colors that appear around each vertex -- were
shown to give rise to an exotic thermodynamic behavior, encompassing lines of
critical points with varying critical exponents~\cite{Castelnovo2004}. The
interplay between interactions and coloring constraints gives rise to a novel
type of dynamical obstruction to equilibration whereby the system freezes into
a polycrystal instead of reaching its ordered ground
state~\cite{Chakraborty2002,Chakraborty2003,Das2003,Castelnovo2004}.

Here we study the effects of nearest-neighbor interactions where the
interaction strength depends on the color of the intervening bond between the
two neighboring vertices (color-dependent interactions). We show that this
leads to an unusually rich phase diagram with different ordered phases
separated by lines, sheets and even three-dimensional regions of critical
points (in parameter space). In particular, the ability to tune the interaction
according to the color of the bond that `carries' it allows us to \emph{break
down} the criticality with central charge $c=2$ of the non-interacting model
and to see it \emph{arise from its originating components}, all the way down to
three Ising $c=\nicefrac{1}{2}$ critical points.

Along the symmetric line, where the three coupling constants are equal, we
observe a line of critical points whose central charge appears (numerically)
to be varying from $c=2$ to $c=\nicefrac{3}{2}$, as first noted in
Ref.~\onlinecite{Castelnovo2006} (Fig.~4). Whilst our work allows to understand
the origin of the value $c=\nicefrac{3}{2}$, the behavior of the system in
between the two points remains a mystery -- plausibly the effect of some
unusually large but not critical correlation length which tricks here (and
nowhere else in the phase diagram) the numerical algorithms into measuring an
incorrect value of the central charge.

The paper is organized as follows. In Sec.~\ref{sec: model}, we introduce the
model and we present a summary of our main results with an overview of the
complete phase diagram of the model. The remaining sections are an account of
the analytical calculations, arguments, and numerics that support our results.
In Sec.~\ref{sec: limiting cases}, we study some useful limits, where
analytical progress can be made by mapping to other known models, and we
compare our findings to numerical results from transfer-matrix calculations. In
Sec.~\ref{sec: the full phase diagram}, we apply the transfer-matrix method
to the rest of the phase diagram of the system. This section contains the most
important results in the paper: the $c=2$ criticality can be viewed as arising
from the merging of $c=1$ free-boson planes of critical points, which in turn
originate each by the merging of two $c=\nicefrac{1}{2}$ planes of critical
points. We argue that the model exhibits a (seemingly supersymmetric)
$c=\nicefrac{3}{2}$ critical point where three Ising $c=\nicefrac{1}{2}$
critical planes merge. And we also observe a region where the central charge
appears to be varying continuously between $\nicefrac{3}{2}$ and $2$ as a
function of the coupling constants -- a likely artifact of an unusually large
but finite correlation length whose origin is yet to be fully understood.
Finally, we draw our conclusions in Sec.~\ref{sec: conclusions}.

Details of the transfer-matrix calculations are given in
App.~\ref{app: TM details}. For completeness, we provide details of Read's
argument for the SU(3)$_1$ CFT description of the non-interacting three
coloring model in App.~\ref{app: Read}.


\section{Model and summary of results}
\label{sec: model and results}

\subsection{Model}
\label{sec: model}

\begin{figure}[b]
\centering
\includegraphics[width=0.85\columnwidth]{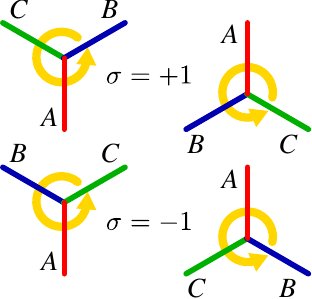}
\caption{
\label{fig: three-color vertices}
(Color online) Allowed vertices in the three coloring model (modulo cyclic
rotations of the colors) and relative values of the chirality spins (namely,
the parity of the color sequence around the vertex in the counterclockwise
direction). [$A$ bonds are red, $B$ bonds are blue, and $C$ bonds are green
throughout the rest of this paper.]
}
\end{figure}

Consider a honeycomb lattice with degrees of freedom living on the bonds and
taking three different values, or colors, $A$, $B$, and $C$, under the
constraint that no two bonds meeting at a vertex can be of the same color.
Each vertex on the lattice must then be in one of the configurations
illustrated in Fig.~\ref{fig: three-color vertices}, up to cyclic rotations
of the colors.

In the non-interacting limit, i.e., from a purely combinatorial perspective,
the model is exactly soluble and it exhibits long range
correlations~\cite{Reshetikhin1991}. As it was elegantly shown by N.~Read at a
kagome workshop in 1992, the discrete $\mathcal{S}_3$ symmetry of the system is
promoted to a continuous SU(3) symmetry of the coarse grained CFT describing
its long wavelength behavior (see App.~\ref{app: Read}).

The model can be alternatively interpreted as a fully-packed loop model with
fugacity $2$. For instance, removing all bonds of a given color, say $C$,
from the lattice yields a fully-packed configuration of closed loops with
alternating coloring $AB$ or $BA$. This in turn allows one to map the three
coloring model onto a two-component height model, which in the non-interacting
limit is sitting precisely at a roughening transition~\cite{Kondev1996-1,*Kondev1996-2}. Note
that one is free to choose a description in terms of $AB$, $BC$, or $CA$ loops:
any three coloring configuration can indeed be seen as the classical
superposition of three coexisting fully-packed loop configurations (strongly
correlated with one another!).

One can introduce chirality spins $\sigma_i = \pm 1$ on the sites of the
honeycomb lattice, the positive sign assigned say to vertices where the colors
appear counterclockwise in an even permutation of the sequence $ABC$ (as
illustrated in Fig.~\ref{fig: three-color vertices}). The three coloring model
can then be mapped onto a constrained Ising system on the sites of a honeycomb
lattice, where each plaquette has either magnetization $0$ or
$\pm 6$.~\cite{Castelnovo2004}

The effect of nearest-neighbor interactions between the spins was studied
in Ref.~\onlinecite{Castelnovo2004} by a combination of numerical and
analytical techniques (see also Ref.~\onlinecite{Castelnovo2006} for further
results). The behavior is surprisingly rich, as highlighted for example by the
fact that weak antiferromagnetic (AFM) interactions do not seem to order the
system, but rather give rise to a mysterious line of critical points with an
apparent `continuously' varying central charge~\cite{Castelnovo2006}. This is
surprising, given the fact that the system is at a roughening transition in the
non-interacting limit, precisely towards the AFM phase, and the addition of AFM
interactions should give a finite mass to the critical modes. Moreover, a
continuously varying central charge is forbidden by Zamolodchikov's c-theorem
in unitary theories; following the general belief that a model with local
constraints and real local energy terms is unitary.

Partly in the attempt to shed light into this unusual behavior, we consider a
generalization of the model in this paper. We assume that the strength of the
nearest-neighbor couplings between the Ising spins depends on the color of
the intervening bond, and the interaction energy is given by,
\begin{equation}
E = -\sum_{\mathclap{l=A,B,C}}J_l~~\sum_{\langle ij\rangle_l}\sigma_i\sigma_j~,
\end{equation}
where $\langle ij \rangle_l$ stands for a pair of neighboring sites $i,j$
connect by a bond of color $l\in\{A,B,C\}$. We investigate the phase diagram of
the system as a function of the reduced couplings $J_l / T$ with $T$ being
temperature, and for convenience of notation we shall directly use the
relabeling
\begin{equation}
J_A \equiv \frac{J_A}{T}, \quad
J_B \equiv \frac{J_B}{T}, \quad
J_C \equiv \frac{J_C}{T}.
\end{equation}
The phase space is spanned by the three real coordinates $J_l \in (-\infty,
\infty)$, which we compactify for convenience to $(-1,+1)$ by introducing the
parameters
\begin{equation}
x_l = \tanh J_l\,, \qquad \text{with } l = A,B,C \, .
\end{equation}

The Boltzmann weight for bond $k$, which connects neighboring vertices $i$ and
$j$, can then be rewritten as
\begin{align}
e^{J_{l_k} \sigma_i \sigma_j}
  =&~e^{J_{l_k}} \delta_{\sigma_i \sigma_j , +1}
	 + e^{-J_{l_k}} \delta_{\sigma_i \sigma_j , -1} \\
  =&~\cosh(J_{l_k}) \left( 1 + x_{l_k} \, \sigma_i \sigma_j \right)\
\, ,
\label{eq:compactified}
\end{align}
where $l_k$ is the color of bond $k$. The factor of $\cosh(J_{l_k})$ can be
neglected, since it contributes a trivial overall factor to the partition
function.

\begin{figure}[tb]
\centering
\includegraphics[width=0.85\columnwidth]{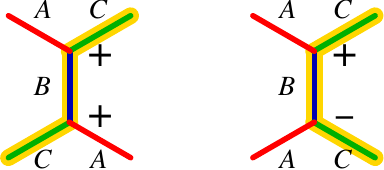}
\centering
\caption{
\label{fig: F vs AF interactions}
(Color online) A FM arrangement of the chirality spins (left) say across a $B$
bond minimizes the local energy for $x_B > 0$ and corresponds to configuring
the adjacent $A$ and $C$ bonds parallel colorwise. This results in `straight'
$AB$ and $BC$ loop segments. Vice versa for an AFM arrangement of spins
(right).
}
\end{figure}

\begin{figure}[tb]
\centering
\includegraphics[width=0.45\columnwidth]{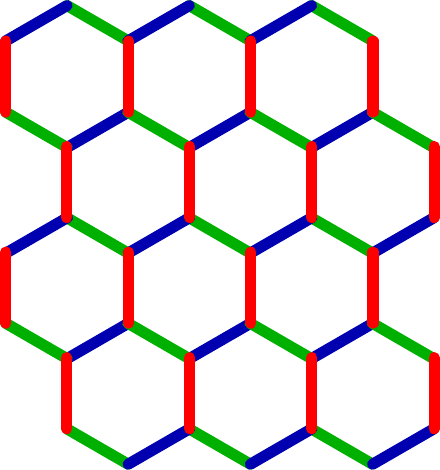}~
\includegraphics[width=0.45\columnwidth]{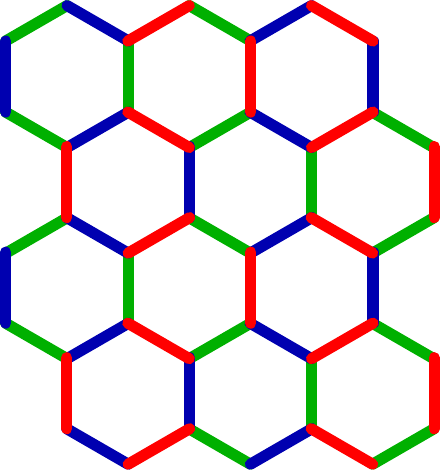}
\caption{
\label{fig: F and AF configs}
(Color online) Examples of fully ferromagnetic (left) and antiferromagnetic
(right) coloring configurations.
}
\end{figure}

A positive (ferromagnetic, FM) coupling $x_l > 0$ favors parallel bonds of the
same color, as illustrated in Fig.~\ref{fig: F vs AF interactions}. As such, it
favors straight loop configurations (i.e., it favors maximal local tilt of the
height mapping in the direction of the bond). Vice versa, a negative
(antiferromagnetic, AFM) coupling $x_l < 0$ favors non-parallel arrangements of
colored bonds and curled loop configurations (i.e., favors locally flat
configurations: if crossing a loop changes the height, crossing an adjacent
loop in the same direction leads to the opposite height change). See
Fig.~\ref{fig: F and AF configs} for examples of fully-FM and fully-AFM
configurations (interactions in dimer-like models that act as loop tension --
equivalently, aligning / anti-aligning terms -- have been considered before on
experimentally relevant grounds, e.g., in Ref.~\onlinecite{Jacobsen2009}).


\subsection{Summary of results}
\label{sec: summary of results}

\begin{figure}[tb]
\centering
\includegraphics[width=0.95\columnwidth]{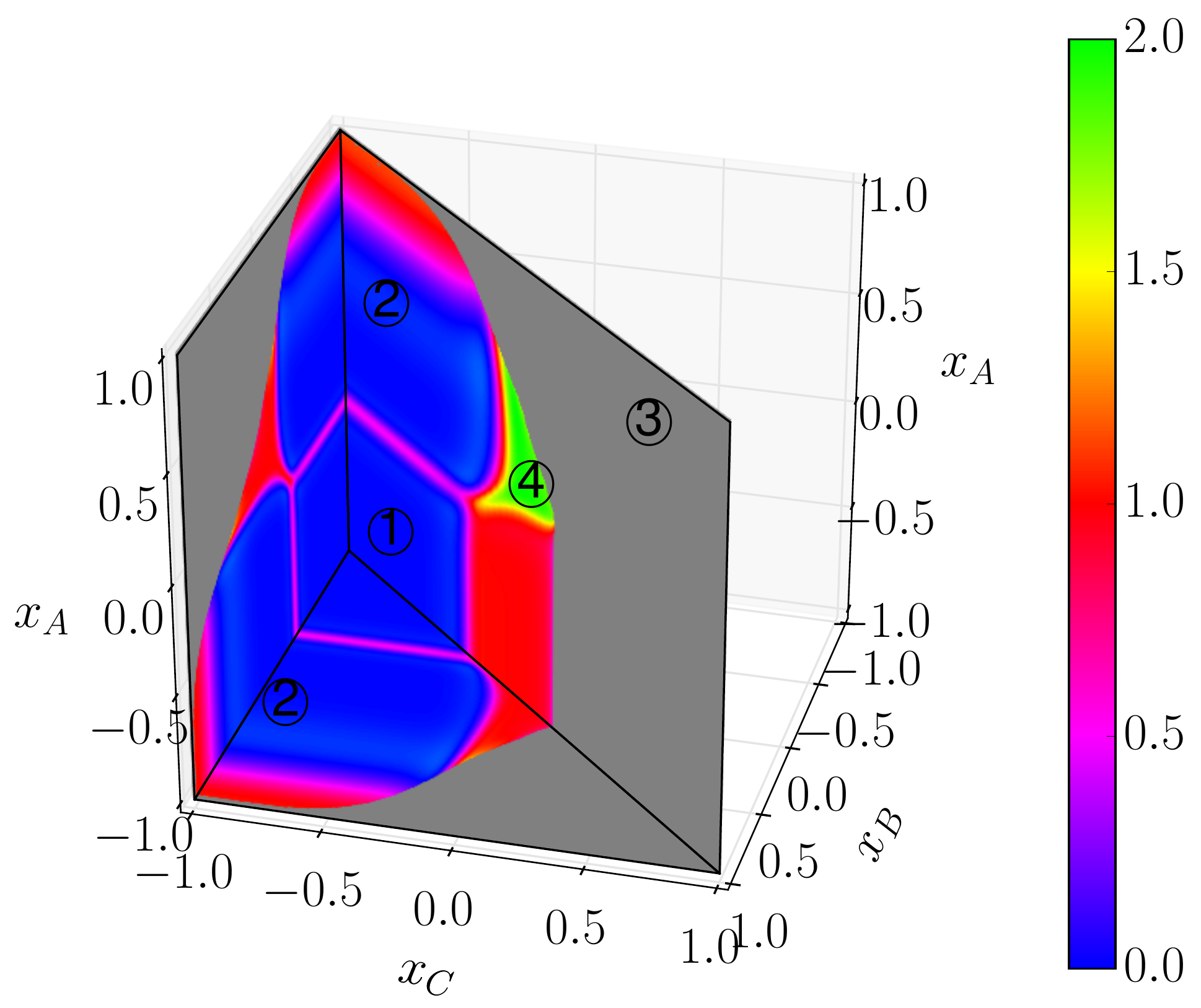}
\caption{
\label{fig:3d-cube-centralcharge}
(Color online)
The two plots of the central charge from Figs.~\ref{fig: JA -inf c Numerics}
and~\ref{fig: JA equ JB Numerics} assembled on a cube. The color scale for the
central charge is shown on the right, and the gray coloring indicates
non-critical regions of propagating color loops. The following regions are
highlighted in this plot: (1) FM-ordered Ising plaquette spin phase with broken
sublattice and $\mathbb{Z}_2$ symmetry, also referred to as
columnar phase; (2) paramagnetic Ising plaquette spin phase with broken
sublattice but restored $\mathbb{Z}_2$ symmetry; (3) non-critical stripe
phases, also referred to as staggered phases; and (4) $c=2$ critical
region containing the non-interacting limit.
}
\end{figure}

For convenience, we summarize here the results of the paper, which make up the
phase diagram of the system illustrated in
Fig.~\ref{fig:3d-cube-centralcharge}.

For large AFM couplings, $x_A, \, x_B, \, x_C \to -1$, the system
orders in a 6-fold degenerate state where all two-color loops are maximally
curled into single hexagons (Fig.~\ref{fig: F and AF configs}, right panel).
We find a cube-shaped region whose phase is continuously connected to this
fully-AFM state (also referred to as the columnar phase). As we explain below,
this phase breaks both lattice translation symmetry and $\mathbb{Z}_2$
symmetry. It can be seen as the FM-ordered phase of three distinct effective
Ising plaquette spins, according to whether we identify the spins with the
orientation of the $AB$, $BC$ or $CA$ loops. Upon increasing either the $x_A$,
$x_B$ or $x_C$ coupling, the system eventually exits the columnar phase into
three distinct paramagnetic phases where the $\mathbb{Z}_2$ symmetry is
restored, but the sublattice symmetry remains broken. The corresponding phase
transition is of the Ising universality class ($c=\nicefrac{1}{2}$, shown in
purple in Fig.~\ref{fig:3d-cube-centralcharge}). These three phases correspond
to disordered Ising phases, but the curled loops live (predominantly) on one of
the three different sublattices of the triangular lattice, face-dual to the
original honeycomb lattice, and they appear to be nowhere continuously
connected with one another in the phase diagram.

When the Ising critical boundaries from different effective descriptions merge
pairwise, our numerics suggest that the $c=\nicefrac{1}{2}$ critical degrees of
freedom fuse to give a $c=1$ free-boson theory at its BKT
transition~\cite{Lecheminant2002,Arlego2003}.
Perhaps even more surprisingly (given that the three Ising descriptions are
in fact not at all independent!), when all three $c=\nicefrac{1}{2}$ merge
at the isotropic $x_A = x_B = x_C$ line, the system exhibits a
$c=\nicefrac{3}{2}$ critical behavior suggestive of a supersymmetric point
(confirming and providing a deeper understanding of, the results already
obtained in Ref.~\onlinecite{Castelnovo2004,Castelnovo2006} along the line
$x_A = x_B = x_C$).

At the phase boundaries where these phases meet \emph{pairwise}, the system
becomes critical with central charge $c=1$ (red regions in
Fig.~\ref{fig:3d-cube-centralcharge}). As the couplings become larger, these
$c=1$ sheets develop into thick `wings' of critical points, of which we
currently lack analytical understanding. The larger the value of the
coupling, the thicker the wings.

\begin{figure}[tb]
\centering
\includegraphics[width=0.45\columnwidth]{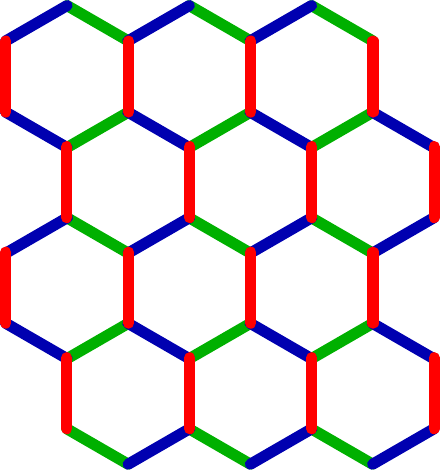}
\caption{
\label{fig: JA = -inf, JC = + inf}
(Color online) Example of a configuration that minimizes the energy of the
system for $x_A \to -1$, $x_C \to +1$.
}
\end{figure}

Beyond the $c=1$ wings, the system enters four different stripe-ordered phases,
which are also referred to as propagating phases, as all two-color loops of
bonds propagate across the entire system and do not form closed loops
(shown in gray in Fig.~\ref{fig:3d-cube-centralcharge}). One of the four
phases is the phase with all bonds being FM aligned, shown in
Fig.~\ref{fig: F and AF configs}, and it is favored when all coupling constants
are strongly FM. The other three phases have either all $A$, all $B$, or all
$C$ bonds AFM aligned, and they appear when respectively $x_A<0$, $x_B<0$, or
$x_C<0$ while the other two coupling constants are strongly FM. An example of
such an ordered configuration is shown in
Fig.~\ref{fig: JA = -inf, JC = + inf}. As discussed in
Sec.~\ref{sec: J_A = -infinity, J_C = +infinity}, we expect these four phases
to be stable not only in the limit of infinite coupling strength, but also in a
finite region of the phase diagram. The transition between the fully-FM and the
three AFM phases is described by a 1D Ising model with nearest-neighbor
interactions whose strength scales with the size of the system (see
Sec.~\ref{sec: J_A,J_B = +infinity}). Hence the transitions between these
different stripe phases is strongly first order, as is the transition between
the stripe phases and the phases that break sublattice symmetry.

In the following sections, we present how these results were obtained using
a combination of analytics and numerics.


\section{Useful limits}
\label{sec: limiting cases}

Let us begin our study of the compactified phase diagram of the model by
considering some informative limiting cases.


\subsection{The $x_A, x_B \to -1$ line}
\label{sec: J_A,J_B = -infinity}

\begin{figure}[tb]
\centering
\includegraphics[width=0.32\columnwidth]{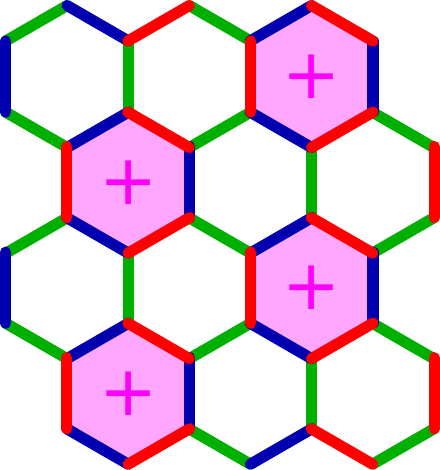}~~
\includegraphics[width=0.32\columnwidth]{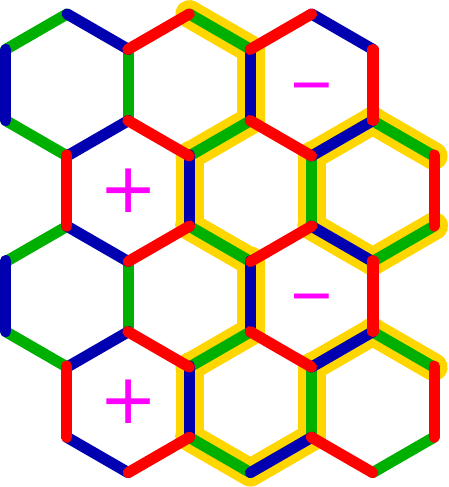}~~
\includegraphics[width=0.32\columnwidth]{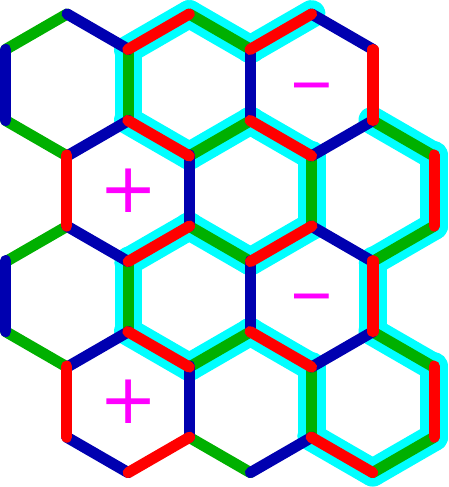}
\caption{
\label{fig: JA=JB=-inf config}
(Color online) Three coloring configurations with $x_A = x_B \to -1$. The
effective Ising degrees of freedom live at the centers of a triangular lattice
formed by one sublattice of the hexagonal plaquettes. (These are \emph{not} to
be confused with the chirality spins introduced earlier.) The left panel shows
one of the two configurations that are selected in the limit $x_C\to-1$. Here
we take the one illustrated as the reference configuration, where all the
effective spins are positive. The middle and right panels show the same color
configuration that differs from the left panel by two effective spins that have
been flipped. Not all hexagonal plaquettes now form two-color loops and longer
loops are present, highlighted in yellow and cyan respectively. By comparing
the panels, one can explicitly see how the $x_C$ interaction in the effective
spin language translates into a nearest-neighbor coupling of strength $-x_C$
between the plaquette spins, where the sign accounts for the fact that $x_C<0$
is now FM.
}
\end{figure}

Consider the limit $x_A = x_B \to -1$, which forces the $AB$ loops to be
maximally curled around single hexagonal plaquettes (see
Fig.~\ref{fig: JA=JB=-inf config}), as a
function of $x_C\in(-1,+1)$ (i.e., along one edge of the compactified phase
diagram, see Fig.~\ref{fig:3d-cube-centralcharge} and
Fig.~\ref{fig: JA -inf c Numerics}). The centers of the hexagonal plaquettes on
the honeycomb lattice form a triangular lattice that is tripartite. Once all
the $AB$ loops form single hexagons, they are bound to occupy exclusively one
of the three sublattices.
In the limit $x_A = x_B \to -1$, the only freedom left
in coloring the system is the orientation of each hexagonal $AB$ loop, say
from $ABABAB$ to $BABABA$, which does not change the sublattice of the dual
triangular lattice occupied by the $AB$ loops. As a result, this phase
breaks lattice translation symmetry into three sectors, depending on
which of the three sublattices the $AB$ loops `condense' on.

Within each sector, all allowed configurations are identified by the
orientations of the $AB$ loops, either $ABABAB$ or $BABABA$. In the limit $x_C
\to -1$, all the loops order with the same orientation, as shown in the left
panel of Fig.~\ref{fig: JA=JB=-inf config}. We can then take one of these two
configurations as our reference and label all others in the same sector using
Ising degrees of freedom living at the centers of the $AB$ hexagonal loops (as
illustrated in Fig.~\ref{fig: JA=JB=-inf config}).
Namely, we can define Ising spins $S_p \in\{+1,-1\}$ at the centers $p$
of the $AB$ plaquettes, where $S_p = +1$ ($S_p = -1$) if the $AB$ hexagon at
$p$ has the same (resp. different) color orientation as in
Fig.~\ref{fig: JA=JB=-inf config} (left panel). What we obtain is a 1-to-1
mapping, modulo the choice of orientation of one $AB$ plaquette, between the
three coloring model in the limit of $x_A = x_B \to -1$ and a triangular
lattice Ising model~\footnote{These effective Ising spins are \emph{not} to be
confused with the chirality spins introduced earlier. Note in particular that,
by construction, a fully-AFM configuration of the chirality spins corresponds
to a fully-FM configuration of the effective spins on the triangular lattice.
Whereas this may seem an annoying source of confusion at the moment, we see
that this choice for the effective spins is in fact of great help in
understanding the behavior of the system along the $x_A = x_B\to-1$ line.}.
With the help of Fig.~\ref{fig: JA=JB=-inf config}, one can verify that a
finite coupling $x_C$ translates into a nearest-neighbor interaction between
the effective spins $S_p$,
\begin{equation}
E = J_C \sum_{\langle p q \rangle} S_p S_q \, .
\end{equation}
Note that $x_C < 0$ is \emph{FM} and $x_C > 0$  is \emph{AFM} (opposite to the
behavior in terms of chirality spins).

Taking advantage of the (exact) mapping in the limit $x_A, x_B \to -1$, we
obtain immediately the behavior of the three coloring model as a function of
$x_C$. Starting from the $x_C \to -1$ limit, the system is in a
FM ordered phase, ending at a second order phase transition of the Ising
universality class (central charge $c=\nicefrac{1}{2}$) at $x_C^* \simeq
-0.26795$ ($J_C^* \simeq -0.27465$)~\cite{Temperley202,*Houtappel1950,
*Wannier1950}. For larger values
of $x_C$, the system enters a disordered phase controlled by the paramagnetic
fixed point $x_C=0$. (Note that only the $\mathbb{Z}_2$ symmetry is restored at
this transition, whereas the lattice translation symmetry remains broken.)

At $x_C=0$, positive and negative $S_p$ spins are distributed randomly with
probability $\nicefrac{1}{2}$ and the model is equivalent to critical site
percolation on the triangular lattice. As a manifestation of the O(1) loop
model, this is part of the dense $c=0$ phase. It is interesting to see how this
single O(1) loop model originates from the three coexisting fully-packed loop
models ($AB$, $BC$ and $CA$) that identify a three coloring configuration
(recall Sec.~\ref{sec: model}). Consider either the ensemble of $BC$ or $CA$
loops on the lattice (we shall see that the two structures give in the end the
same coarse grained loop model). When three neighboring effective spins $S_p$
have the same sign, the corresponding $AB$ hexagons have all the same
orientation and the hexagonal plaquette in the middle of the three spins must
have alternating coloring, either $BC$ or $CA$ (see
Fig.~\ref{fig: JA=JB=-inf config}). By construction, this plaquette sits in the
bulk of a domain of the effective Ising model. Since in site percolation one is
interested in the domain boundaries, we shall remove this single $BC$ or $CA$
hexagon from the corresponding $BC$ or $CA$ loop ensemble, without losing any
information. After repeating this operation throughout the lattice, we are left
with $BC$ and $CA$ loop configurations that are no longer fully packed. One can
further verify that for every loop in the former, there is a unique loop in the
latter having precisely the same backbone (see the yellow and cyan shaded loops
in Fig.~\ref{fig: JA=JB=-inf config}) and vice versa. In other words, the two
loop configurations are effectively identical and they trace the domain walls
in the $S_p$ Ising model. These are nothing but the conventional O(1) domain
wall loops in the loop gas construction by Nienhuis~\cite{Nienhuis_loopgas} and
the critical (site percolation) behavior can be directly inferred from them.

When $x_C > 0$, the interactions between the effective spins become AFM and the
model is frustrated. The disordered phase survives for any finite positive
$x_C$, and $x_C = 0$ is the fixed point for the entire $x_C^* < x_C < +1$
basin. Spatial correlations diverge again in the limit $x_C \to +1$, where the
system is equivalent to the zero-temperature limit of the classical Ising AFM
on the triangular lattice. This fully-frustrated system can be mapped onto a
dimer model on the dual honeycomb lattice (which is \emph{not} the same as the
original lattice of the three coloring model), whose long wavelength behavior
is captured by a free-boson CFT with central charge $c=1$.

Notice that the fully-frustrated triangular Ising model is only obtained if the
limit $x_C \to +1$ is taken \emph{after} $x_A,x_B \to -1$. In
Sec.~\ref{sec: J_A = -infinity, J_C = +infinity} we shall see how an entirely
different behavior arises if for instance we take the limit $x_A \to -1$ first,
then $x_C \to +1$ and then $x_B \to -1$. We postpone the discussion of this
issue of order of limits to Sec~\ref{sec: (+,-,-) corners}.

Notice that the system is generically symmetric under any permutation of the
colors and the considerations in this section extend straightforwardly to the
lines $x_B = x_C \to -1$ and $x_A = x_C \to -1$ in the phase diagram. Similarly
for results presented in the following sections.

The results of this section are reported in the left panel of
Fig.~\ref{fig: JA -inf c Numerics} in {\textbf{\color{mymagenta}magenta}}.


\subsection{The $x_A \to -1$, $x_C \to +1$ line}
\label{sec: J_A = -infinity, J_C = +infinity}

Let us consider then the other distinct edge of the back plane $x_A \to -1$ in
parameter space (Fig.~\ref{fig:3d-cube-centralcharge}, and also
Fig.~\ref{fig: JA -inf c Numerics}), namely the limit $x_A \to -1$, $x_C \to
+1$ (equivalently, $x_A \to -1$, $x_B \to +1$).

It is convenient to start by setting $x_B = 0$. The energy of the system is
then minimized by configurations where all $A$ bonds are AFM ordered and all
$C$ bonds are FM ordered. These conditions can be satisfied throughout the
lattice without frustration, and $6$ configurations are selected, related by
symmetry to the one shown in Fig.~\ref{fig: JA = -inf, JC = + inf}.

In these configurations, all $BC$ loops are straight (i.e., the chirality spins
are FM ordered along the loops) and there are perfect AFM correlations across
the loops. Notice that all of the configurations are necessarily FM ordered
across the $B$ bonds, even in the absence of $x_B$ interactions: the very same
phase is obtained upon taking $x_B \to +1$ and $x_C \to +1$, and then taking
$x_A \to -1$. In the chirality spin language, this constitutes a
\emph{stripe phase}.

Clearly, these remain the lowest (free) energy configurations for $0 < x_B <
+1$. They in fact remain the lowest energy configurations in the entire region
of phase space where $x_A<0$ and $x_B,x_C>0$. However, they have no entropy and
one would need to assess whether they are stable in presence of thermal
fluctuations. In analogy with previously studied dimer/loop models with
tension,~\cite{Castelnovo2007,Jacobsen2009} it is reasonable to envision that
the stripe phase ($x_A \to -1$, $x_B \geq 0$, and $x_C \to +1$) survives in a
finite 3D region of the phase diagram and is not lost as soon as the reduced
couplings become finite. Indeed, one can (qualitatively) view $x_A<0$ and
$x_B,x_C>0$ as tension terms in a fully-packed $BC$ loop model on the honeycomb
lattice. The latter is expected to enter the staggered phase where all the
loops are completely straight at some finite value of the reduced tension.

The case of a $-1 < x_B < 0$ tends to destabilize the stripe phase. However,
this is unlikely to occur abruptly and the phase should survive a finite extent
into the phase diagram for large but finite values of $x_C$.

Notice the peculiar entropic behavior of these ordered configurations. As one
can directly verify in Fig.~\ref{fig: JA = -inf, JC = + inf}, they do not allow
finite energy fluctuations. All the two-color loops wind around the system.
Consequently, the smallest re-arrangement that is obtained by exchanging the
colors along one of the loops has an energy cost that scales with the linear
size $L$ of the system, whereas the entropic gain scales only as $\ln(L)$. All
fluctuations about these configurations are infinitely suppressed in the
thermodynamic limit: they form vanishing entropy basins in the free energy
landscape of the three coloring model. We expect the system to enter or exit
this phase via a first-order transition.

The results of this section are reported in the left panel of
Fig.~\ref{fig: JA -inf c Numerics} in \textbf{\color{myyellow}yellow}.


\subsection{Non-commuting order of limits in the $(-1,-1,+1)$ corner}
\label{sec: (+,-,-) corners}

Notice that the behavior of the system near each of the three corners $(-1,-1,
+1)$, $(-1,+1,-1)$, and $(+1,-1,-1)$ in parameter space depends on the
direction of approach. Consider for instance the case $x_A = -1$. If you first
take $x_C \to +1$ and then $x_B \to -1$, the system is locked into one of the 6
stripe configurations discussed in
Sec.~\ref{sec: J_A = -infinity, J_C = +infinity}. On the other hand, if you
first take $x_B \to -\infty$ and then $x_C \to +1$ you enter the
fully-frustrated limit (dual to a triangular Ising AFM) discussed in
Sec.~\ref{sec: J_A,J_B = -infinity}.

The two phases are not continuously connected and the order of limits matters.
In the neighborhood of $(-1,-1,+1)$, the frustrated phase has lower energy and
finite entropy, and we expect it to dominate in parameter space. This is indeed
reflected in the numerical results in Sec.~\ref{sec: J_A = -infinity}.


\subsection{The $x_A \to -1$, $x_B = x_C$ line}
\label{sec: J_A = -infinity, exact diagonal}

Along the diagonal $x_B=x_C$ of the $x_A \to -1$ plane (see
Fig.~\ref{fig: JA -inf c Numerics}), it is convenient to describe the system as
a $BC$ loop model. One can verify that it is always possible to color
\emph{any} fully-packed loop configuration on the honeycomb lattice so that the
chirality spins are \emph{AFM correlated across all of the bonds that are not
part of the loop configuration}~\footnote{
The reader familiar with the height mapping discussed in
Ref.~\onlinecite{Kondev1996-1,*Kondev1996-2} may recall that any two-color loop configurations
in the three coloring model can be seen as equal height contours in one of the
two height components (for an appropriate choice of the height mapping
vectors). AFM correlations between loops imply that if the height goes up
crossing one loop in a given direction, then it must go down crossing the
adjacent loop in the same direction. A few drawings should suffice to convince
oneself that it is always possible to consistently construct a surface
compatible with the contour loops, given this recipe. On the other hand,
FM interactions require the same height change to occur crossing adjacent loops
in the same direction and this condition cannot be always satisfied by a single
valued surface (consider for instance a hexagon with three adjacent loops). The
only loop configurations where FM correlations are not frustrated need to have
all the loops parallel to each other and therefore winding around the system
(there are clearly $\sim 2^L$ such configurations).}. Indeed, there are
precisely two such coloring patterns per loop configuration.  Therefore, if the
non-interacting three coloring model can be viewed as a fully-packed $BC$ loop
model with fugacity $2$, taking the limit $x_A \to -1$ simply locks the $BC$
loop coloring with one another and reduces the fugacity from $2$ to $1$. It has
no effect on the choice of loop covering.

As a result, we obtain a fully-packed loop model on the honeycomb lattice with
fugactiy $1$. The couplings $x_B = x_C$ provide a tension term
acting along the loops. This model was studied by Jacobsen and Alet for $x_B =
x_C < 0$~\cite{Jacobsen2009}. The case of both positive and negative tension,
albeit on the square rather than the honeycomb lattice, was studied in
Refs.~\onlinecite{Alet2006,Papanikolaou2007,Castelnovo2007}.

The tensionless limit $x_B = x_C = 0$ is equivalent to a dimer model on the
honeycomb lattice, whose long wavelength correlators are the same as in a
$c=1$ free-boson CFT. Couplings $x_B = x_C > 0$ induce a tension term that
favors straight loops. This leads to a line of critical points where $c=1$
survives up to a first-order phase transition to the `staggered' phase at a
finite value of the coupling. Similarly on the AFM side of the interactions
$x_B = x_C < 0$, except that the transition to the `columnar' phase is of the
Beresinskii-Kosterlitz-Thouless kind.

The results of this section are reported in the left panel of
Fig.~\ref{fig: JA -inf c Numerics} in {\textbf{\color{myturquois}turquois}}.


\subsection{Numerical results on the $x_A \to -1$ plane}
\label{sec: J_A = -infinity}

\begin{figure*}[tb]
\centering
\includegraphics[height=5.3cm]{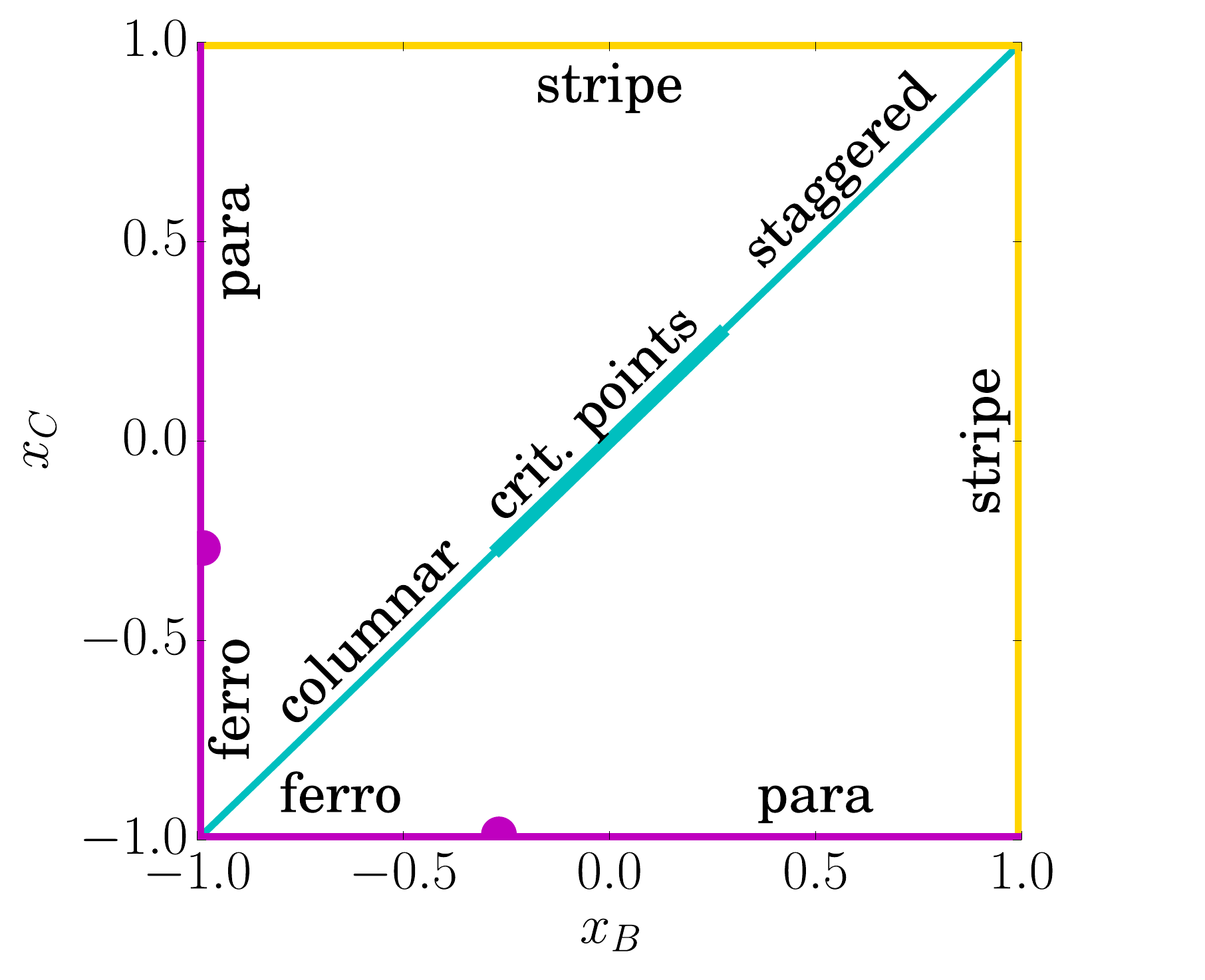}~
\includegraphics[height=5.3cm]{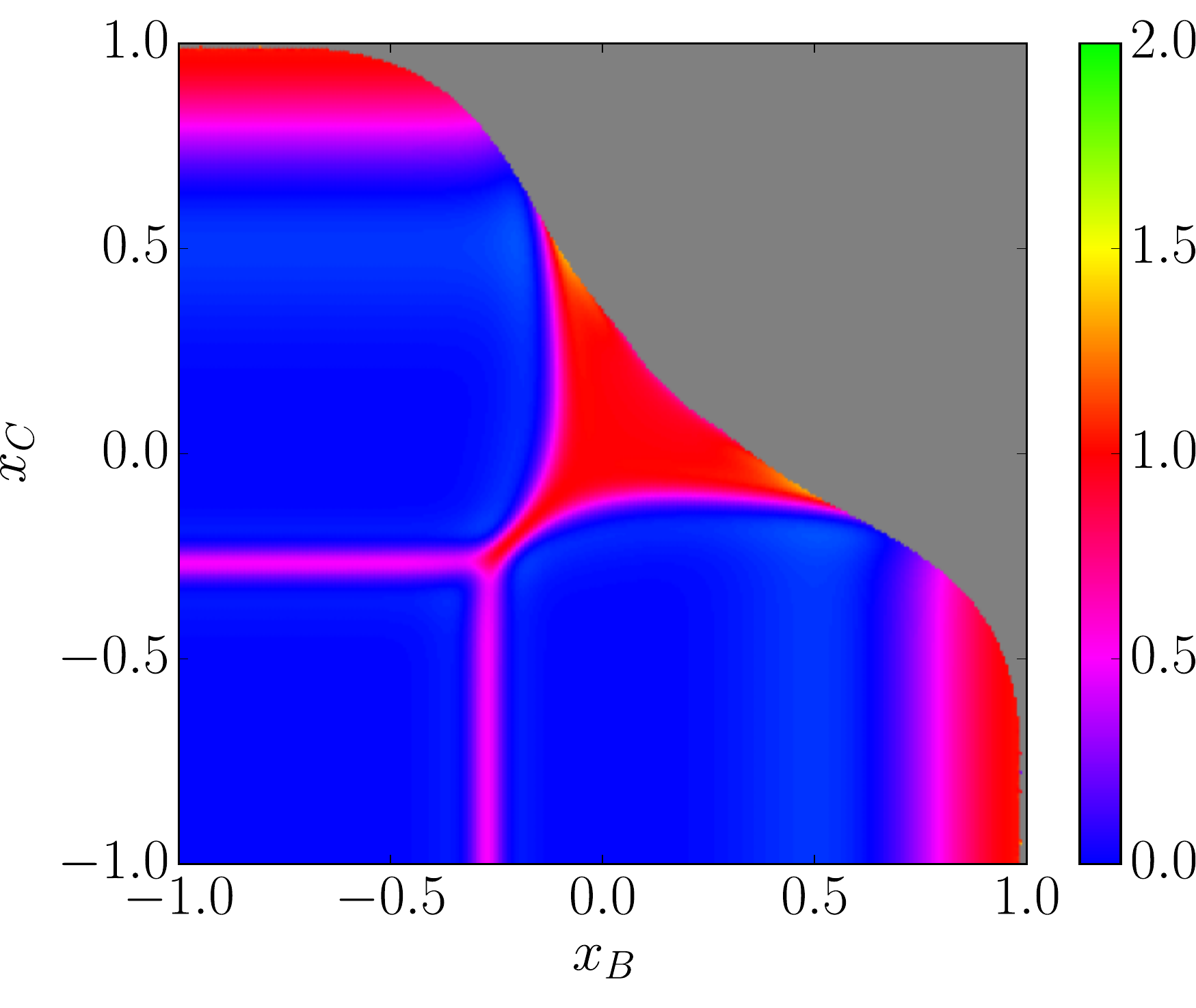}~
\includegraphics[height=5.3cm]{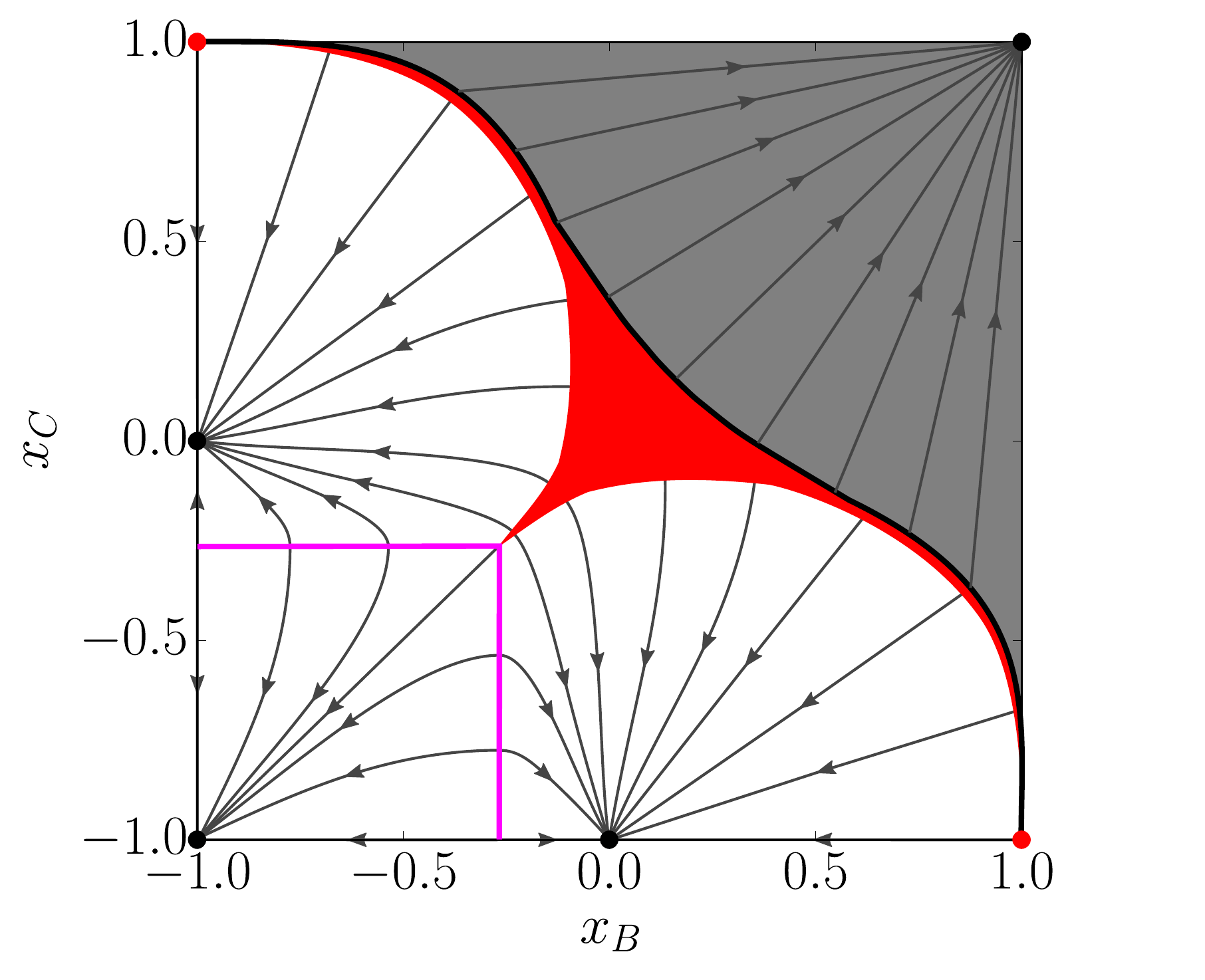}
\caption{
\label{fig: JA -inf c Numerics}
(Color online) Left: Schematic illustration of the limits of the model
considered in
Sec.~\ref{sec: J_A,J_B = -infinity}-\ref{sec: J_A = -infinity, exact diagonal}
on the $x_A \to -1$ plane. Middle: Intensity plot of the central charge
$c(x_B,x_C)$ from numerical simulations on the $x_A\to-1$ plane. The color
scale for the central charge values is on the right. Lattices of size
$N = 18,24,30$ were used to obtain this data. Right: Schematic renormalization
group flow lines for the $x_A \to -1$ plane. Note that RG flow lines have to
protrude from critical lines or regions in a perpendicular way.
}
\end{figure*}

In order to complete the phase diagram of the system on the $x_A \to -1$ plane,
away from the limiting cases considered above, we calculate the central charge
as a function of $x_B$ and $x_C$ using a transfer-matrix approach. This allows
to identify the phase-transition boundaries between the massive phases
(where $c=0$), and provides also an initial characterization of the
critical behavior.

The transfer matrix for adjacent row colorings is constructed for a three
coloring configuration on a cylinder, and the free energy in the thermodynamic
limit of infinite cylinder length can be obtained from the largest eigenvalue
of the matrix. Computing the reduced free-energy density $f$ (i.e., measured
per temperature and unit area) for different values of the system size along
the circumference $L$ of the cylinder, and using the relation~\cite{Bloete1986,
*Affleck1986}
\begin{equation}
f(L) = f_0 - \frac{\pi c}{6 L^2},
\end{equation}
the central charge $c$ can be determined from the finite-size scaling of the
free energy, where $f_0$ is the infinite-size free energy of the system. When
constructing the transfer matrix, we find that its eigenspaces decompose
into sectors of fixed numbers of bonds of the three colors on the rows winding
around the cylinder. We focus only on the sector with equal number of
bonds of each color, as this sector contains all the coloring configurations
with no two-color loops propagating around the cylinder or along its length.
Fluctuations of the latter configurations are suppressed in the
thermodynamic limit,
and hence the corresponding transfer-matrix sectors will be relevant only for
phases of the system that are dominated by zero-entropy configurations, similar
to the ones discussed in Sec.~\ref{sec: J_A = -infinity, J_C = +infinity}. We
expect those phases to be non-critical, and to be connected via first-order
phase transitions.

We give a detailed description of the construction of the transfer matrix, the
exploitation of various symmetries, and the finite-size scaling in
App.~\ref{app: TM details}.

We plot the numerical results for the central charge $c$ as a function of $x_B$
and $x_C$ on the $x_A \to -1$ plane in Fig.~\ref{fig: JA -inf c Numerics}
(middle), where the portion of the diagram colored in gray corresponds to the
propagating phase of the system. These results are in good agreement with the
analytical arguments provided in
Sec.~\ref{sec: J_A,J_B = -infinity}-\ref{sec: J_A = -infinity, exact diagonal},
which are indicated in Fig.~\ref{fig: JA -inf c Numerics} (left).

A first-order phase transition line separates the propagating (gray) phases
from the non-propagating (colored) phases, and is in good agreement with the
first-order transition of the loop model with tension discussed in
Sec.~\ref{sec: J_A = -infinity, exact diagonal}, as well as with the predicted
ground state for the $x_A \to -1$, $x_C \to +1$ line discussed in
Sec.~\ref{sec: J_A = -infinity, J_C = +infinity}.

The locations of the $c=\nicefrac{1}{2}$ ridges for $x_C \to -1$ and $x_B \to
-1$ are in excellent agreement with the value $x_B^*, \, x_C^* = -0.26795$,
that the mapping to the triangular lattice Ising models predicts (see
Sec.\ref{sec: J_A,J_B = -infinity}). In the corners of the plane where
$x_B \to +1$, or $x_C \to -1$ respectively, the numerics indicate $c=1$, which
is also in good agreement with the dimer models on a honeycomb lattice dual to
the triangular lattice of the plaquette centers that is discussed in
Sec.~\ref{sec: J_A,J_B = -infinity}. The $c=\nicefrac{1}{2}$ transition lines
and the $c=1$ critical lines are surprisingly parallel to the coordinate axes,
suggesting that the softening of the constraint that allowed the mapping to an
effective Ising model does not actually alter the free energy of the latter.

The Ising $c=\nicefrac{1}{2}$ lines merge into a $c=1$ critical point when they
meet at the $x_B=x_C$ symmetry line. This appears to coincide with the end of
the line of $c=1$ critical points of the $x_B=x_C$ loop model, where it
undergoes a BKT transition to the `columnar' phase and we can report this point
to be at $x_B = x_B^*$ and $x_C = x_C^*$. A discussion of the merging of two
free Majorana fermion CFTs into a free boson CFT at a BKT transition can be
found in Refs.~\onlinecite{Lecheminant2002,Arlego2003}.

While the $c=\nicefrac{1}{2}$ ridges are consistent within numerical accuracy
with 1D lines of critical points, the numerical results around the $c=1$
$x_B=x_C$ critical line suggest that it extends into a finite 2D critical
region upon increasing $x_B$ and $x_C$.

We note that the numerical results cannot conclusively rule out the possibility
that the $c=1$ wings that originate from the fully-packed loop model on the
$x_B = x_C$ line is connected to the $c=1$ region originating from the
fully-frustrated AFM on the triangular lattice in the corners of the phase
diagram. This issue is discussed in further detail in
App.~\ref{app:fitting-finite-size}.

To check that the different critical behaviours observed in this model are
mutually consistent, we present a plausible sketch of renormalization group
(RG) flow lines on the
$x_A \to -1$ plane in Fig.~\ref{fig: JA -inf c Numerics} (right).
For concreteness, we consider only the case where the $c = 1$ critical
regions are continuously connected.


\subsection{The $x_B,x_C \to +1$ line}
\label{sec: J_A,J_B = +infinity}

\begin{figure}[b]
\centering
\includegraphics[width=0.46\columnwidth]{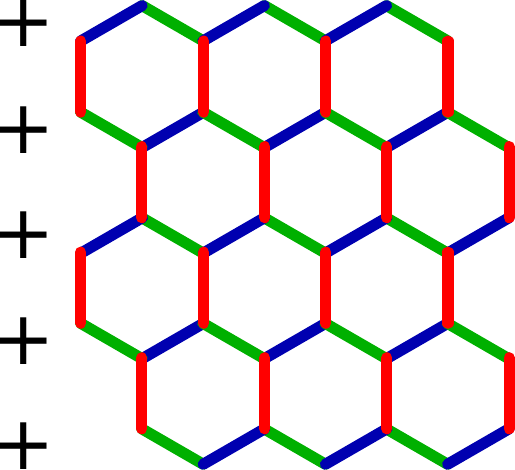}
\hfill
\includegraphics[width=0.46\columnwidth]{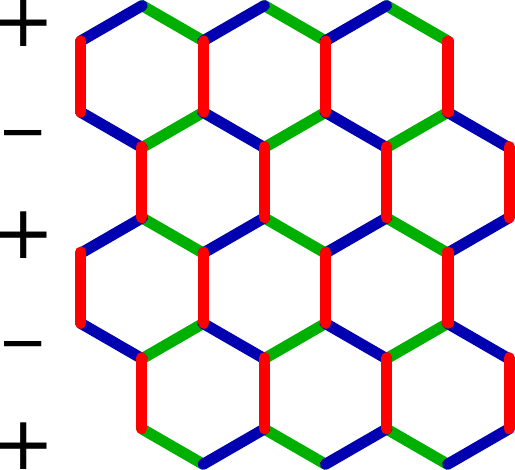}
\caption{
\label{fig: JA,JB = +inf}
(Color online) Example of configurations that minimize the energy of the system
for $x_B,x_C \to +1$. The left panel is favored by $x_A > 0$, the right panel
by $x_A < 0$.
}
\end{figure}

In order to investigate what happens as we leave the three `back' planes in the
phase diagram in Fig.~\ref{fig:3d-cube-centralcharge}, let us again consider
first some limiting cases. We start here by discussing the line $x_B,x_C \to
+1$ (i.e., one of the remaining independent edges of the compactified phase
diagram), where the $BC$ loops are all straight. The $x_A$ interaction couples
the coloring patterns in neighboring $BC$ loops, so that two adjacent loops
arranged as in the left panel of Fig.~\ref{fig: JA,JB = +inf} have energy
$\propto -J_A L$, whilst two adjacent loops arranged as in the right panel of
Fig.~\ref{fig: JA,JB = +inf} have energy $\propto +J_A L$, where $L$ is the
linear size of the system (in the horizontal direction in
Fig.~\ref{fig: JA,JB = +inf}) and we recall that $J_A = \operatorname{artanh}
(x_A)$. We assume here for simplicity a square system of size $L \times L$.

The system behaves as a classical 1D Ising chain of length $L$, where the Ising
degrees of freedom correspond to the coloring sequence of each $BC$ loop
(either $BCBCBC\ldots$ or $CBCBCB\ldots$). In this language, $J_A$ gives rise
to a nearest neighbor interaction of effective strength $\propto J_A L/2$.
Customarily, a 1D Ising model with short range interactions does not order at
finite values of the couplings. However, in this case the induced coupling is
proportional to the length $L$ of the system, and hence it orders FM (AFM) for
any $x_A > 0$ ($x_A < 0$). Indeed, neglecting the constant contribution due to
$x_B$ and $x_C$, the free energy of the system is given by
\begin{align}
\begin{cases}
F \sim - J_A L^2 \, , & J_A < 0 \, , \\
F \sim \ln\left( 2^{L} \right) = \ln(2) L \, , & J_A = 0 \, , \\
F \sim + J_A L^2 \, , & J_A > 0 \, .
\end{cases}
\end{align}

In analogy with the discussion in
Sec.~\ref{sec: J_A = -infinity, J_C = +infinity}, both phases ($x_A < 0$ and
$x_A > 0$) are zero entropy basins of the free energy. Any transition into and
out of these phases is therefore expected to be strongly first order.


\subsection{The $x_B=x_C=0$ line}
\label{sec: J_A = J_B = 0}

\begin{figure}[b]
\centering
\includegraphics[width=0.95\columnwidth]{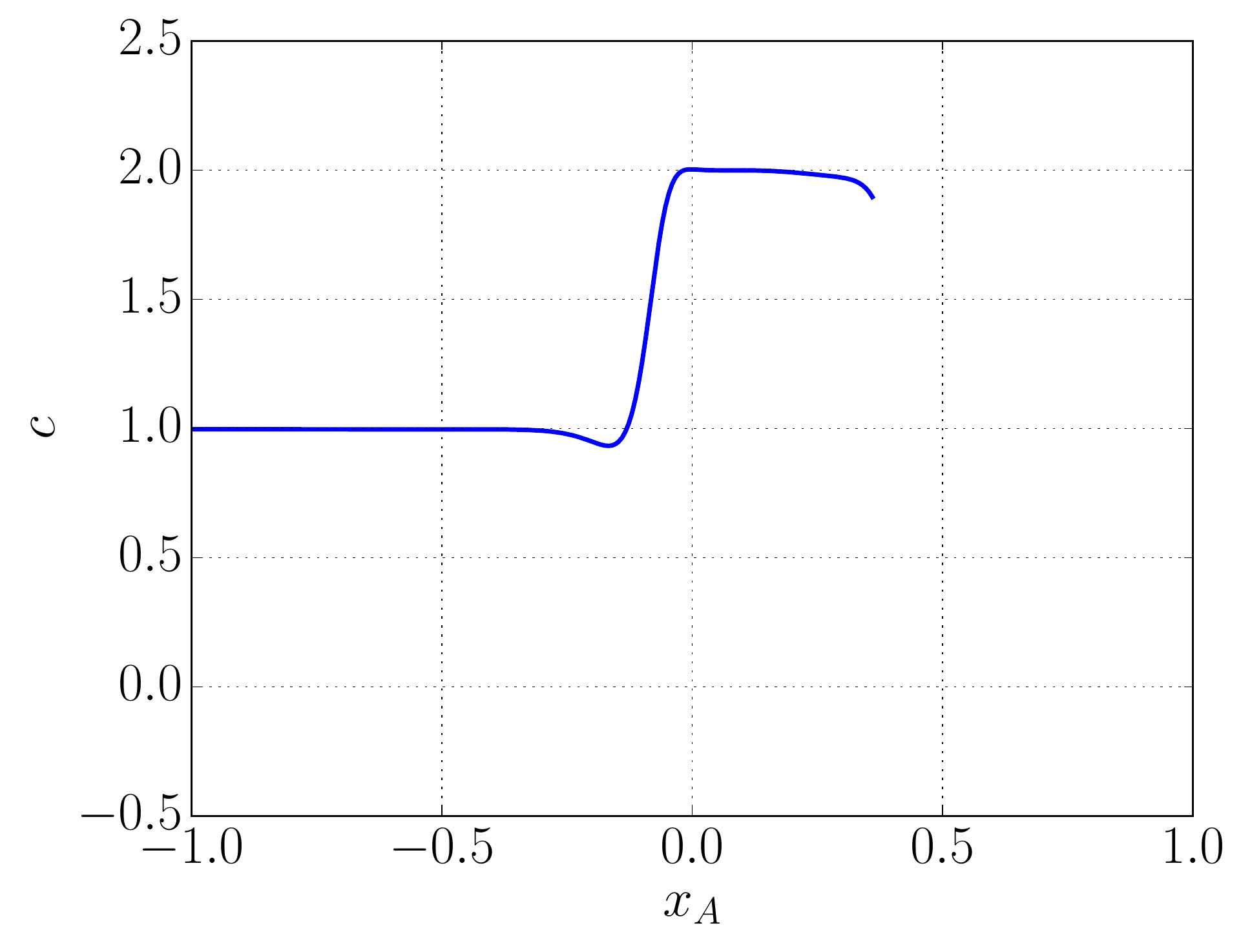}
\caption{
\label{fig: JA = JB = 0}
Behavior of the central charge of the system along the line $x_B=x_C=0$, as a
function of $x_A$.
}
\label{fig:xA=xB=0}
\end{figure}

Consider then the line where $x_B=x_C=0$, as a function of $x_A$. Similarly to
the discussion in Sec.~\ref{sec: J_A = -infinity, exact diagonal}, the behavior
of the system is most readily understood if we interpret it as a fully-packed
loop model of $BC$ loops. The fact that $x_B=x_C=0$ means that there is no loop
tension and the remaining interaction $x_A$ couples the coloring patterns
($BCBCBC\ldots$ or $CBCBCB\ldots$) on adjacent loops.

In the limit $x_A \to -1$, the loops lock into an AFM pattern (i.e., the
chirality spins have perfect AFM correlations \emph{between loops}). As already
mentioned in Sec.~\ref{sec: J_A = -infinity, exact diagonal}, one can verify
that such correlations are never frustrated. That is, it is always possible to
minimize each and every $x_A$ interaction by choosing an appropriate coloring
pattern for \emph{any} chosen loop configuration~\cite{Note3}.
In this limit, one recovers the fully-packed loop model on the honeycomb
lattice with fugacity $1$, which is critical with central charge $c=1$.

The non-interacting $x_A = 0$ point is a fully-packed loop model with fugacity
$2$ (i.e., we are free to choose either the $BCBCBC\ldots$ or the
$CBCBCB\ldots$ coloring for each loop). This is again critical, with central
charge $c=2$. One expects the system to simply transition from one critical
theory to the other as a function of $x_A < 0$ and this is indeed confirmed
numerically (see Fig.~\ref{fig:xA=xB=0}). However, a rigorous analytical
argument to capture the transition is not readily available.

The case $x_A > 0$ is intrinsically different, since in the limit $x_A \to +1$
the system becomes `frustrated': most $BC$ loop configurations do not allow for
a coloring that produces FM correlations across all $A$ bonds. A large positive
$x_A$ progressively selects configurations that are compatible with FM order
between loops. This is a subextensive set of all configurations (there are
$\sim 2^L$ of them), characterized by having all $BC$ loops winding and
parallel to each other (notice that they need not be straight, hence their
number scales with $2^L$). They correspond clearly to a non-critical, massive
phase ($c=0$).

Again, there is no available analytical argument to understand the transition
from the $c=2$ critical theory at $x_A=0$ and the massive phase that obtains
for $x_A \to +1$. Numerical transfer-matrix results below indicate that it
happens at a finite value of $x_A > 0$ and the essentially staggered nature of
the ordered phase suggests that the transition ought to be first order.


\subsection{The $x_A=0$, $x_B = x_C$ line}
\label{sec: other limits}

\begin{figure}[b]
\centering
\includegraphics[width=0.95\columnwidth]{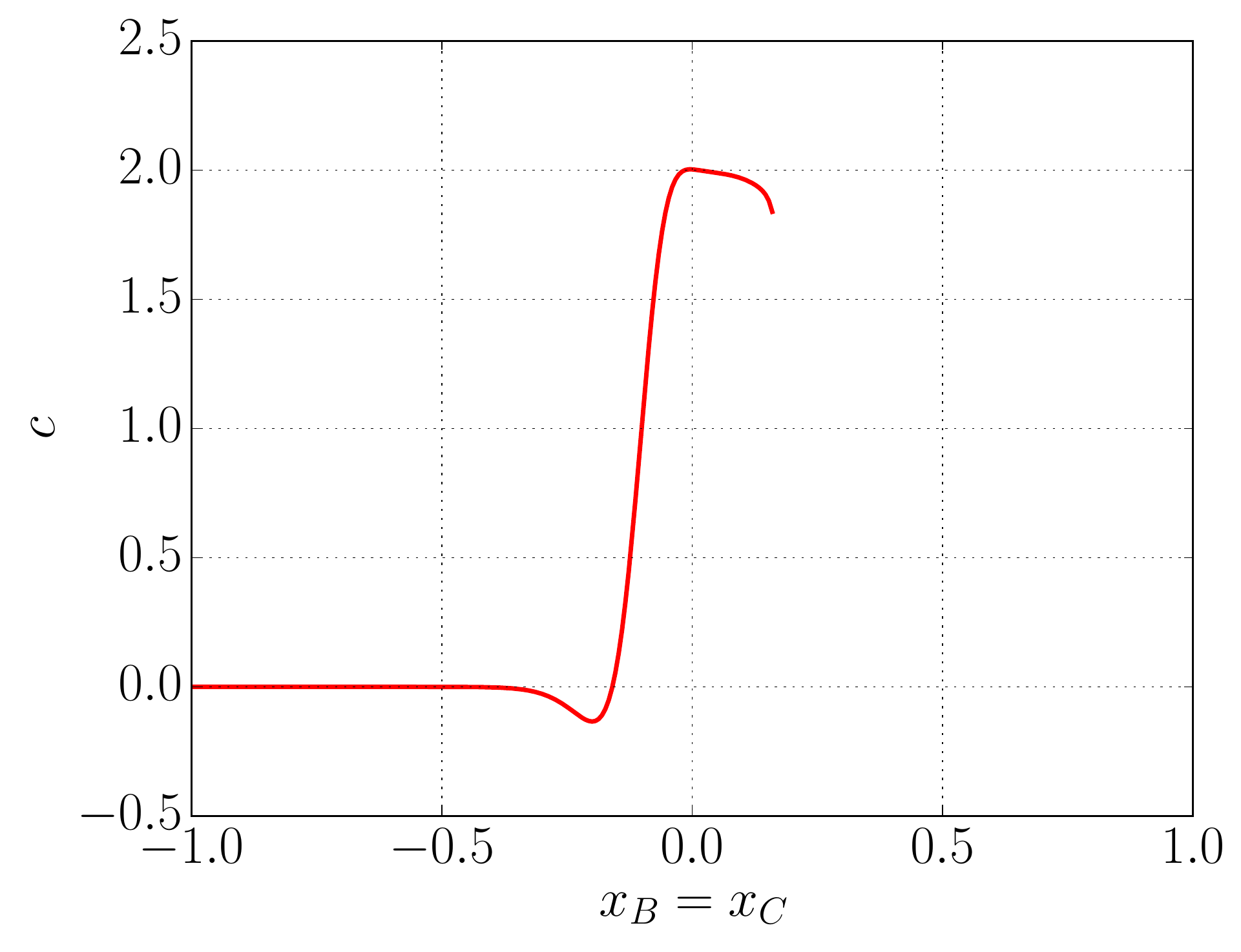}
\caption{
\label{fig: JA = 0, JB = JC}
Behaviour of the central charge of the system along the line $x_A=0$,
$x_B=x_C=x^\prime$.
}
\end{figure}

Once again, along the line $x_A=0$ as a function of $x_B=x_C=x^\prime$,
it is convenient to view the system as a fully-packed loop model of
$BC$ loops. The loops are not interacting across $A$ bonds. On the other hand,
the coupling $x^\prime$ translates into a tension term along the loops
(compare with the discussion in
Sec.~\ref{sec: J_A = -infinity, exact diagonal}).

In the limit $x^\prime \to +1$, the loops are forced to be as straight as
possible, with a residual $2^L$ degeneracy due to the fact that each loop can
be colored either $BCBCBC\ldots$ or $CBCBCB\ldots$, irrespectively of its
neighbors (see Sec.~\ref{sec: J_A,J_B = +infinity}). Vice versa, for $x^\prime
\to -1$ the loops are curled into single hexagons, with the same residual
degeneracy as that of an Ising paramagnet on the triangular lattice, as
discussed in Sec.~\ref{sec: J_A,J_B = -infinity}.

The non-interacting $x^\prime = 0$ point is a fully-packed loop model with
fugacity $2$, which is critical with central charge $c=2$. In analogy with the
behavior of the case with fugacity $1$ as a function of tension (see
Sec.~\ref{sec: J_A = -infinity, exact diagonal} and
Refs.~\onlinecite{Alet2006,Castelnovo2007,Papanikolaou2007,Jacobsen2009}), one
expects the system to exhibit a line of $c=2$ critical points that terminates
into a BKT transition towards the columnar phase and into a first-order
transition towards the staggered phase. This is indeed consistent with the
behavior of the central charge that is obtained from transfer-matrix
calculations, illustrated in Fig.~\ref{fig: JA = 0, JB = JC}.


\subsection{The isotropic line: $x_A = x_B = x_C$}
\label{sec: J_A = J_B = J_C}

\begin{figure}[b]
\centering
\includegraphics[width=0.95\columnwidth]{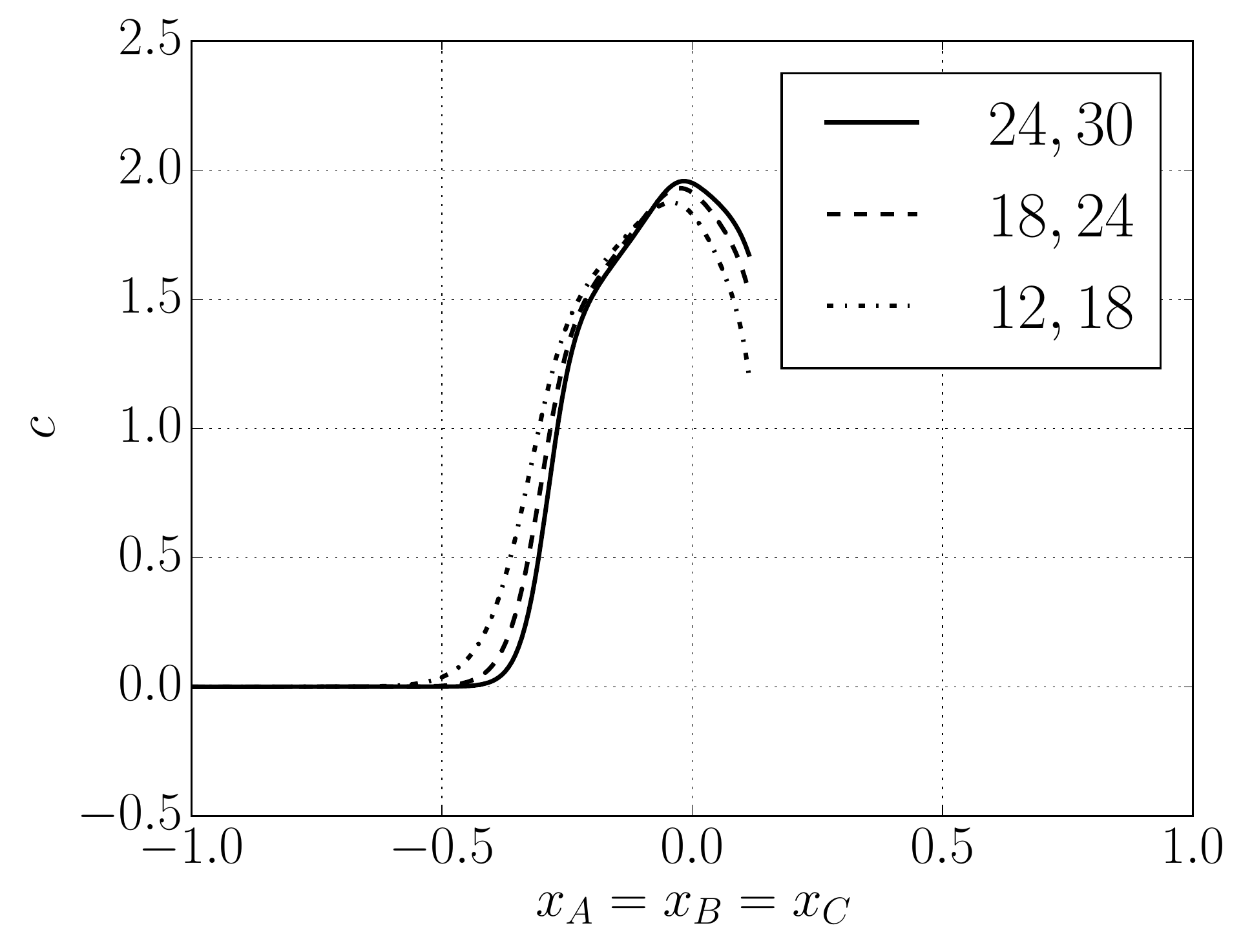}
\caption{
\label{fig: JA = JB = JC}
Behavior of the central charge of the system along the line $x_A=x_B=x_C$. The
central charge is obtained from finite-size fittings of the free energy for
systems with system size $N$ equal to 1) $30$ and $24$, 2) $24$ and $18$, and
3) $18$ and $12$.
Notice the significantly reduced finite size change of the central charge
between $c=2$ and $c=\nicefrac{3}{2}$ (on the negative side of the
horizontal axis), with respect to the behaviour outside this range.}
\end{figure}

We finally consider the phase diagram in the isotropic limit of this model
($x_A = x_B = x_C \equiv x$), which was already studied in detail in
Refs.~\onlinecite{Castelnovo2004,Castelnovo2006},
using both numerical (transfer-matrix and Monte-Carlo
simulations) as well as analytical (cluster mean-field) techniques. We briefly
summarize it here for completeness.

Consistently with the discussion above, the system enters a `columnar' phase
for $x \to -1$. Each two-color loop is maximally curved around single hexagonal
plaquettes (AFM state). Vice versa, for $x \to 1$ the system enters a
`staggered' phase where all the loops are maximally straight and wind around
the system (FM state).

In between these two phases, the system exhibits a line of critical points,
ending in a continuous transition towards the columnar phase and in a strongly
first order transition towards the staggered phase.
As observed previously~\cite{Castelnovo2006}, we find a remarkably small
change with system size in the central charge between the $c=2$ and
$c=\nicefrac{3}{2}$ points, in contrast to the far more substantial drift
outside this range (see Fig.~\ref{fig: JA = JB = JC}).
We shall discuss this behavior in greater detail in
Sec.~\ref{sec: the full phase diagram} and in Sec.~\ref{sec: conclusions}.


\section{The full phase diagram}
\label{sec: the full phase diagram}

\begin{figure}[b]
\centering
\includegraphics[width=0.95\columnwidth]{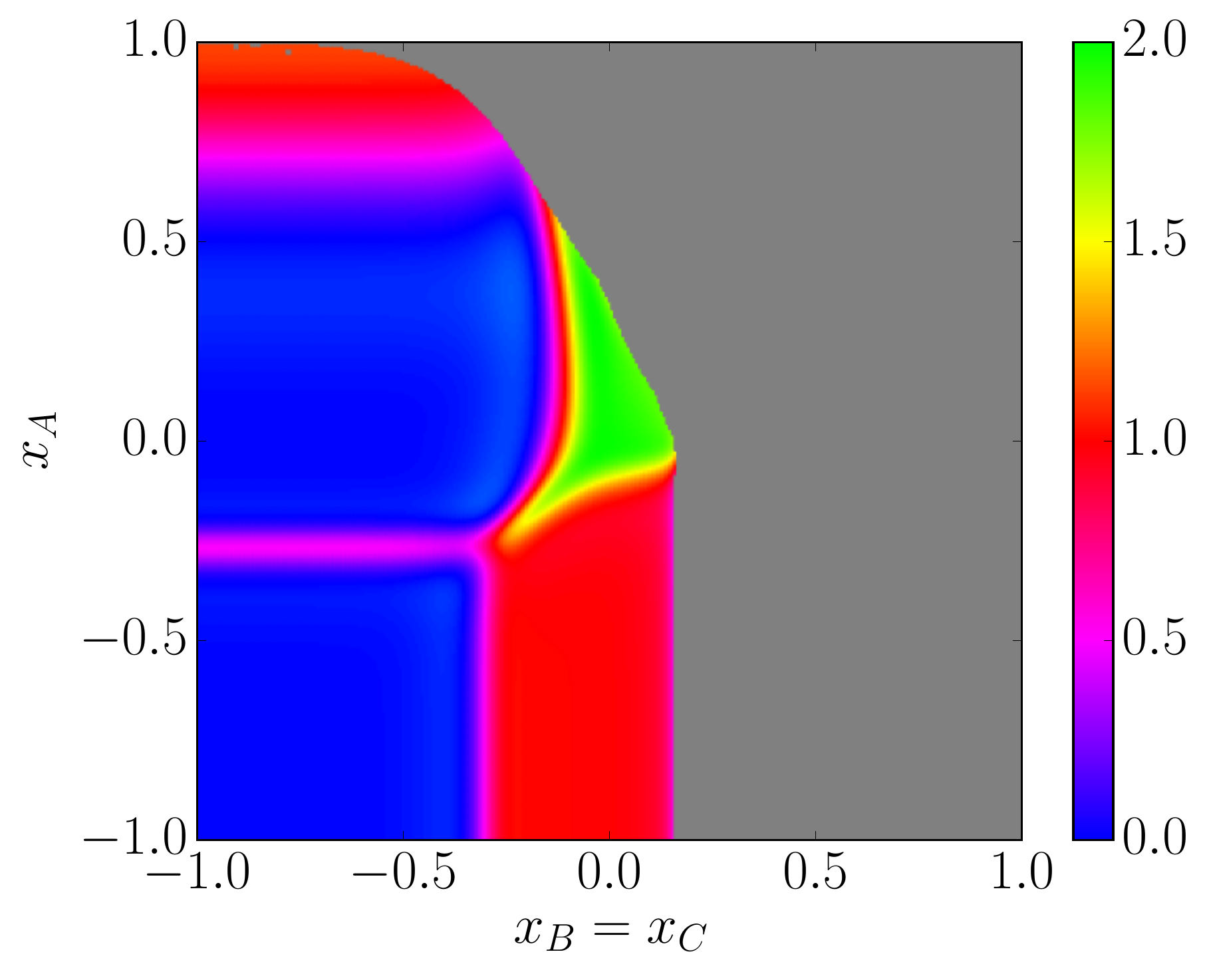}
\caption{
\label{fig: JA equ JB Numerics}
(Color online) Plot of $c(x_A, x_B=x_C)$ from numerical simulation on the
$x_B=x_C$ plane. Lattices of size $N = 12, 18, 24$ were used to obtain this
data. The color scale for the central charge values is on the right.
}
\end{figure}

\begin{figure}[t]
\centering
\includegraphics[width=0.45\columnwidth]{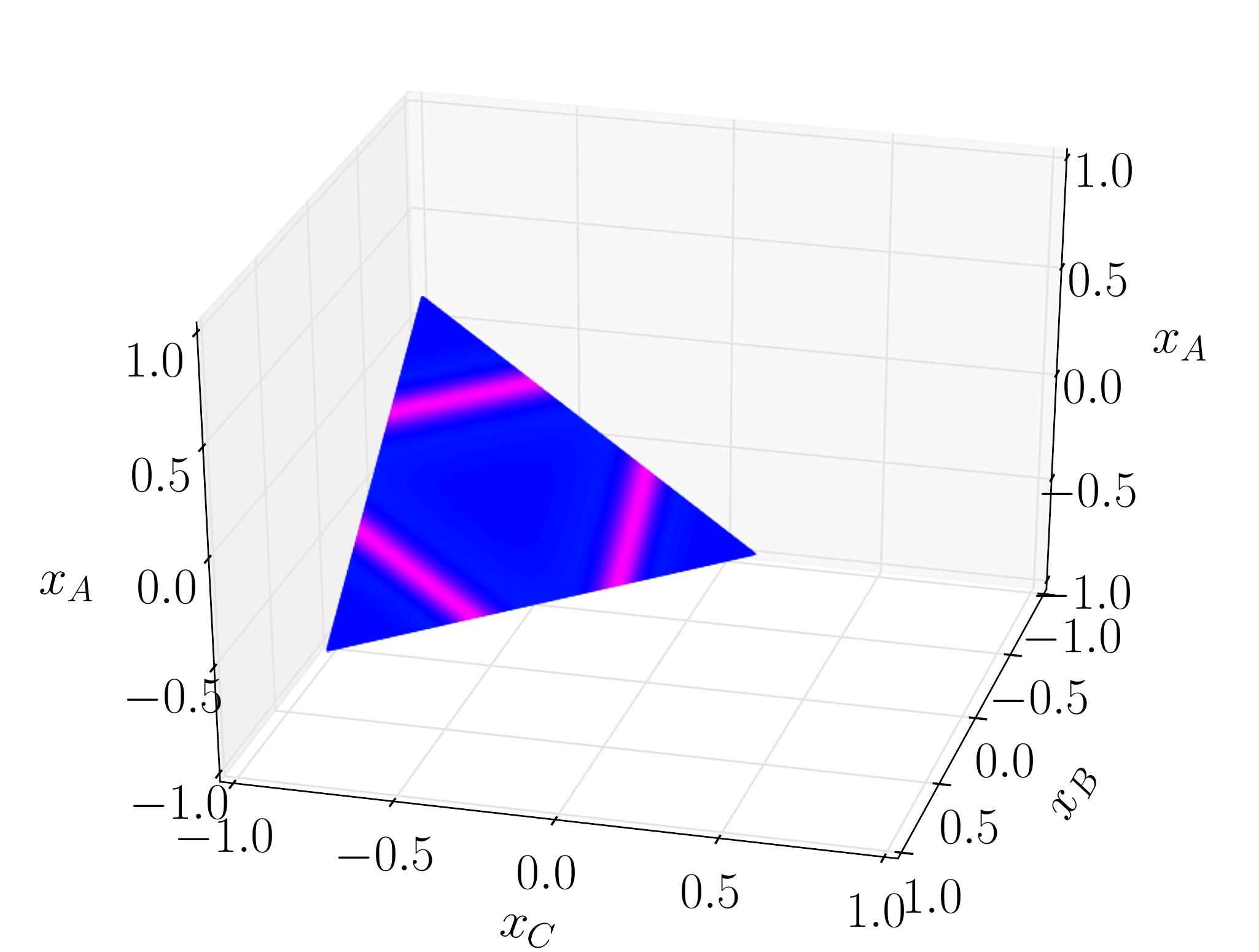}
\includegraphics[width=0.45\columnwidth]{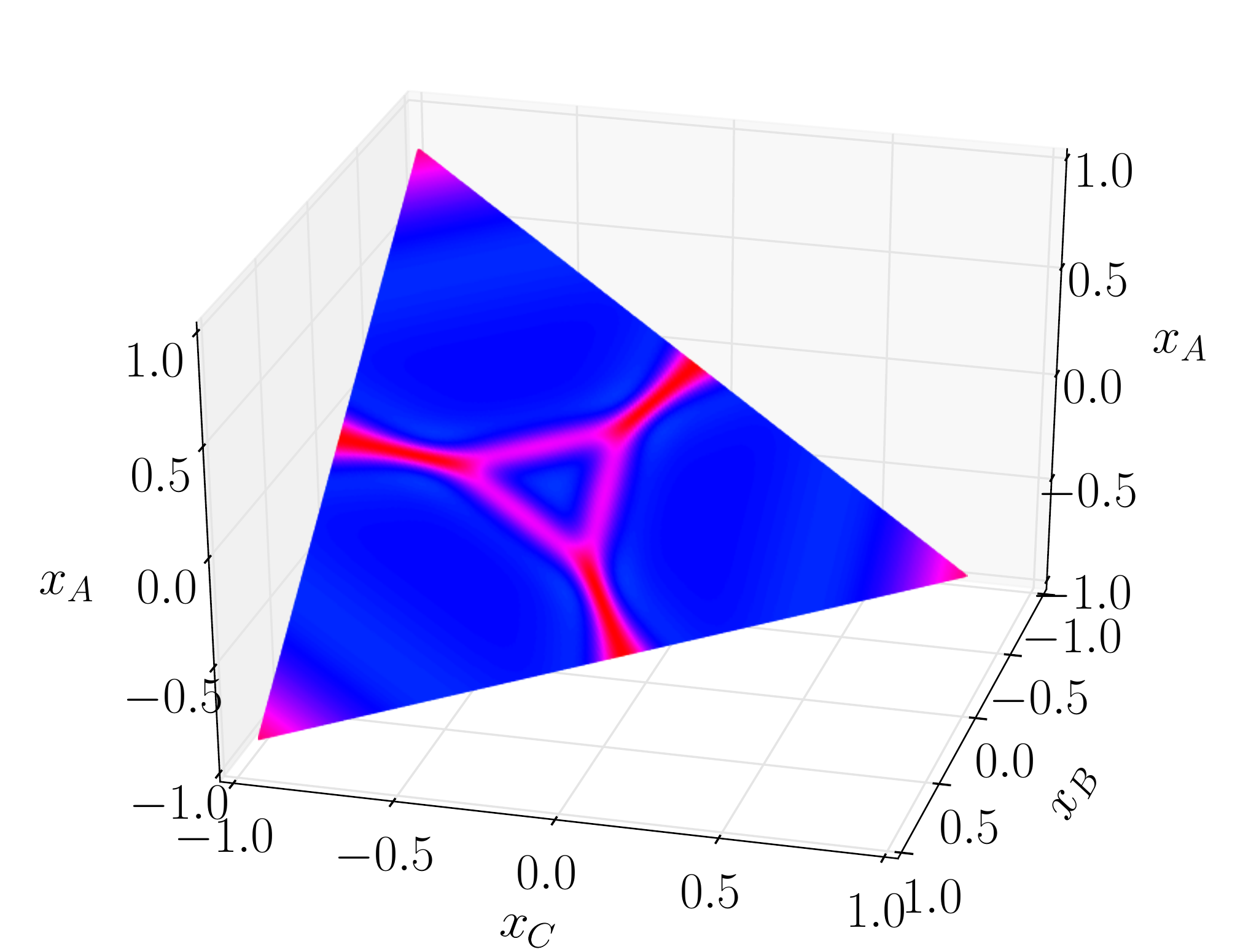}
\includegraphics[width=\columnwidth]{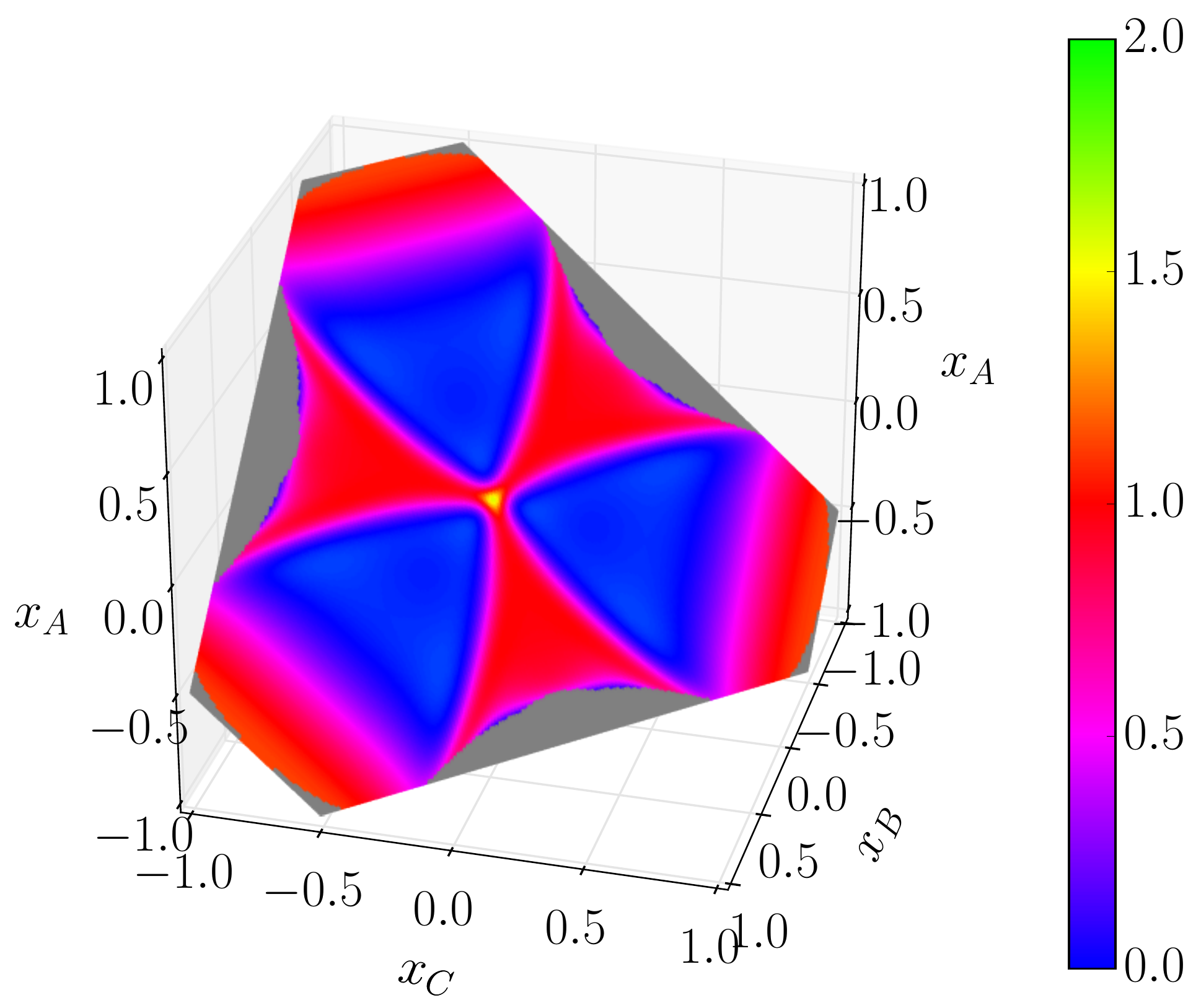}
\includegraphics[width=0.45\columnwidth]{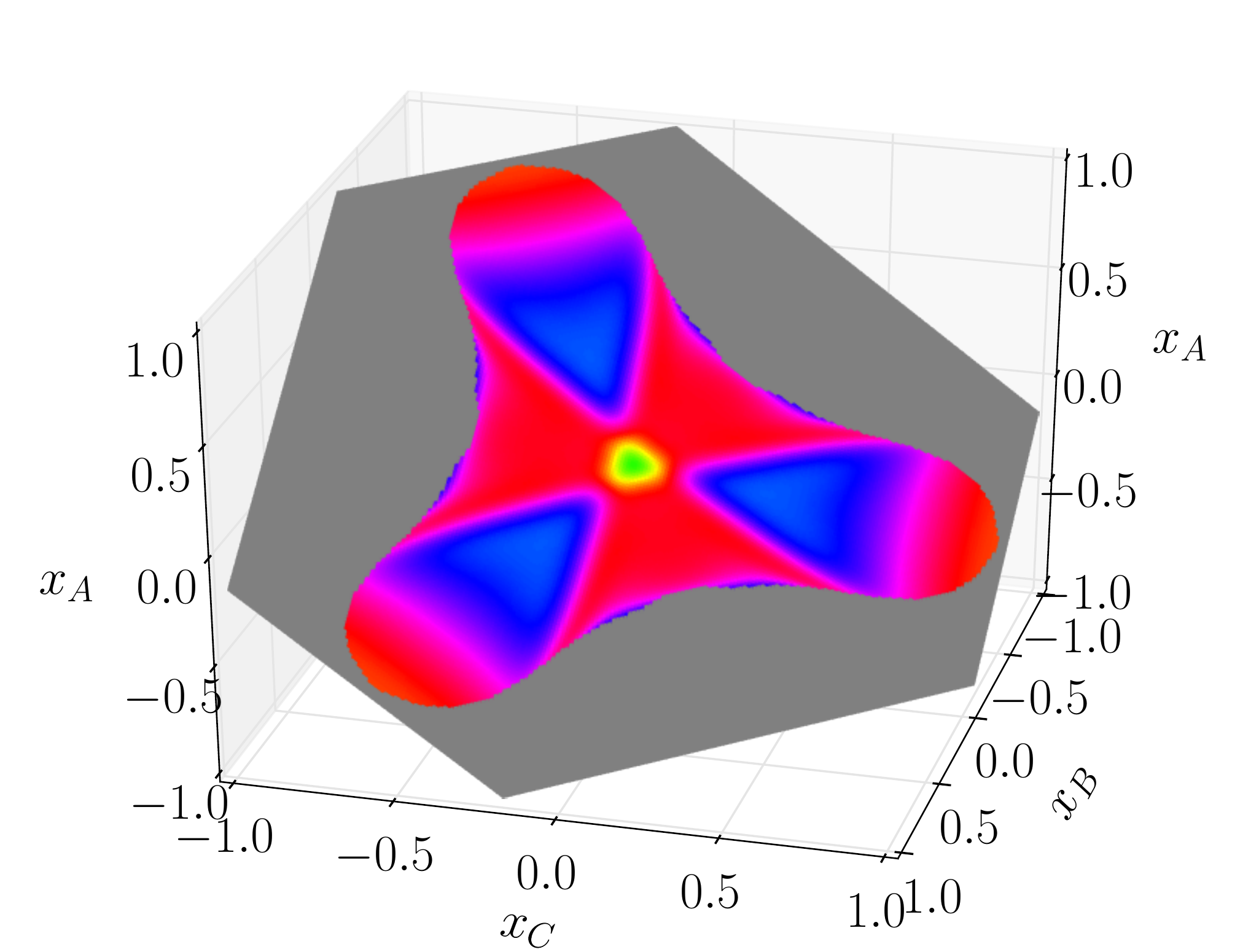}
\includegraphics[width=0.45\columnwidth]{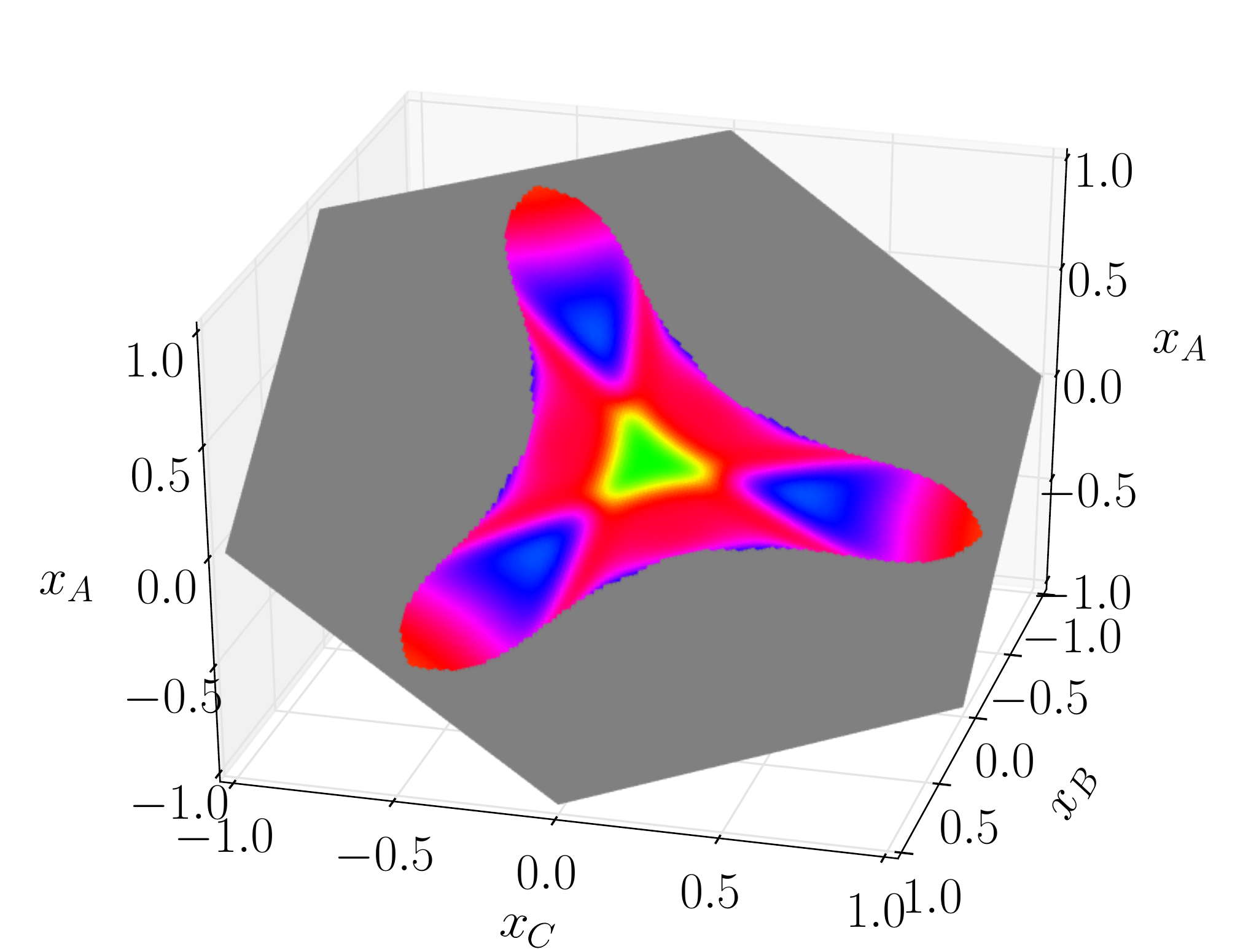}
\caption{
\label{fig: Tri_Plots}
(Color online) Plots of $c$ from numerical simulation on planes perpendicular
to the $x_A=x_B=x_C$ line. Planes with distance $-1.1$, $-0.7$, $-0.3$, $-0.1$,
and $0.0$ from the origin are shown. Lattices of size $N = 12,18,24$ were used
to obtain this data. The color scale for the central charge values is on the
right.
}
\end{figure}

To complete the picture, we use the
numerical transfer-matrix approach to study other 2D slices through the 3D
phase diagram. In Fig.~\ref{fig: JA equ JB Numerics}, we show the central
charge values on the plane $(x_A, x_B=x_C)$ that interpolates between the
$\mathbb{Z}_2$ Ising effective description (left vertical axis) and the
fully-packed loop model with tension (bottom horizontal axis). This allows us
to understand that the $c=1$ free boson line on the horizontal axis, extends
into a region for finite (negative) values of $x_A$.

We also note that the $c=\nicefrac{1}{2}$ lines seen in
Fig.~\ref{fig: JA -inf c Numerics} extend in fact to 2D $c=\nicefrac{1}{2}$
planar sheets forming three adjacent faces of the approximate cube containing
the columnar ordered phase. The horizontal $c=\nicefrac{1}{2}$ line in
Fig.~\ref{fig: JA equ JB Numerics} runs along the diagonal of one such
cubic face. When the cubic faces meet, two distinct $c=\nicefrac{1}{2}$
critical degrees of freedom merge into a $c=1$ BKT edge, bordering the $c=1$
sheets discussed above. This is precisely what happens along the vertical
left-most edge of the $c=1$ sheet in Fig.~\ref{fig: JA equ JB Numerics}.

Comparing these results with cuts across planes perpendicular to the isotropic
$x_A=x_B=x_C$ axis, shown in Fig.~\ref{fig: Tri_Plots}, we see that when the
three $c=1$ sheets merge at a point $x_A=x_B=x_C \simeq -0.25$,
they give rise to a
region with apparent `smoothly increasing' central charge. This region starts
from $c=\nicefrac{3}{2}$ at the point where the three $\mathbb{Z}_2$ Ising
critical points merge -- suggestive of supersymmetric properties -- and
increases up to $c=2$ at approximately $x = 0$.

For positive values of the couplings, the $c=1$ sheets develop into thick
wing-shaped regions (see Fig.~\ref{fig: JA -inf c Numerics} in
Sec.~\ref{sec: J_A = -infinity}). From Fig.~\ref{fig: JA equ JB Numerics} and
Fig.~\ref{fig: Tri_Plots} we learn that, as these wings merge
in the bulk of the phase diagram, they give rise to an extended 3D region
with central charge $c=2$, inclusive of the exactly solvable
non-interacting point $x_A = x_B = x_C = 0$.

Once again, the gray regions in Fig.~\ref{fig: JA equ JB Numerics} and
Fig.~\ref{fig: Tri_Plots} correspond to stripe and staggered (i.e.,
FM) phases, as discussed in Sec.~\ref{sec: J_A = -infinity}. The
numerical results along the edges of the gray regions are consistent with the
conjectured first-order nature of the transition.


\section{Conclusions}
\label{sec: conclusions}

We already summarized our main results and the phase diagram of the system in
Sec.~\ref{sec: summary of results}. The most controversial and interesting
behavior occurs when the three sublattice-translation symmetry-broken phases
meet along the isotropic line $x_A=x_B=x_C \lesssim 0$. Here, the three
different $c=1$ free boson sheets meet but their critical behavior cannot add
up (as is the case instead for the $c=\nicefrac{1}{2}$ sheets meeting at the $c
= \nicefrac{3}{2}$ point) because they are effective descriptions of a model
that can host at most two free bosonic degrees of freedom and not three.

The result is a puzzling (numerical) central charge, which starts from
$c=\nicefrac{3}{2}$ at a finite $x < 0$ and appears to increase continuously --
to the best of our finite size scaling -- up to $c=2$
(see Fig.~\ref{fig: JA = JB = JC}). However, the $c$-theorem forbids a
continuously varying central charge along a line of fixed points in a unitary
CFT.

In the study of the full phase diagram of the interacting three coloring model,
we have seen how one can sometimes map the system onto a fully-packed loop
model with different values of the fugacity. Interestingly, if the fugacity is
varied continuously between $1$ and $2$ in such a model, unitarity is lost and
the system remains critical in between the two unitary limits, with a
continuously varying central charge between $1$ and $2$~\cite{Bloete1994}. One
may therefore
wonder whether the interacting three coloring model may in fact be non-unitary.
However, the fact that the constraints are local and the interactions are real
and local suggest otherwise, and we cannot offer a convincing argument in favor
of non-unitarity. An alternative intriguing conjecture is that the RG flow
lines of the unitary three coloring model along the $x_A=x_B=x_C$ line may run
close to those of a non-unitary fully-packed loop model with continuously
varying fugacity, thence exhibiting its scaling behavior up to very large
length scales (beyond which the true scaling of the unitary model would be
reveled -- length scales which are unfortunately not accessible using our
numerical transfer-matrix approach). What the true unitary behavior of the
system is in between the $c=\nicefrac{3}{2}$ and $c=2$ points remains therefore
elusive: one possible scenario is that the central charge remains constant at
$\nicefrac{3}{2}$ up to the $x_A=x_B=x_C \sim 0$ point, where it increases to
$2$; the difference between the two CFTs is a massive vs. massless Majorana
fermion, which may be responsible for an unusually long correlation length that
is mistakenly picked up by the finite-size scaling as an effective contribution
to the central charge. However, one can clearly envisage many alternative
scenarios and further work is needed to fully elucidate this conundrum.

The software to generate the numerical values for the central charge, the data
obtained from it for the cases studied and plotted in this paper, and the
scripts to generate those plots from the data are available
online~\cite{cambridge_apollo}.


\section*{Acknowledgments}
This work spurred from a collaboration with C.~Mudry and C.~Chamon, and the
authors are deeply indebted to them, for several of the ideas regarding the
isotropic limit of the model developed from those early discussions. Since its
inception in 2005, several collaborators joined and left this project, with
noteworthy contributions from F.~Trusselet and P.~Pujol. We are particularly
grateful to Nick Read for sharing his private notes written at a kagome
workshop in 1992, and for taking the time to discuss them with us in detail.
Part of those notes are reproduced with his permission in App.~B of this
manuscript.  We are indebted to J.~Cardy for several insightful discussions and
for pointing out the $c$-theorem argument against a continuously varying
central charge in Sec.~\ref{sec: conclusions}. Furthermore, P.~Verpoort thanks
V.~Jouffrey, M.J.~Rutter, and B.~Andrews for helpful discussions concerning the
numerical calculations. This work was supported in part by Engineering and
Physical Sciences Research Council (EPSRC) Grant
No.~GR/R83712/01 and by EPSRC Postdoctoral Research Fellowship EP/G049394/1
(C.~Castelnovo), and by EPSRC Grant No. EP/D070643/1 (JJHS). P. Verpoort
acknowledges funding by the Studienstiftung des deutschen Volkes. Statement of
compliance with the EPSRC policy framework on research data: this publication
reports theoretical work that does not require supporting research data.


\appendix

\section{Details of the transfer-matrix calculations}
\label{app: TM details}

\begin{figure}[b]
\centering
\includegraphics[width=0.95\columnwidth]{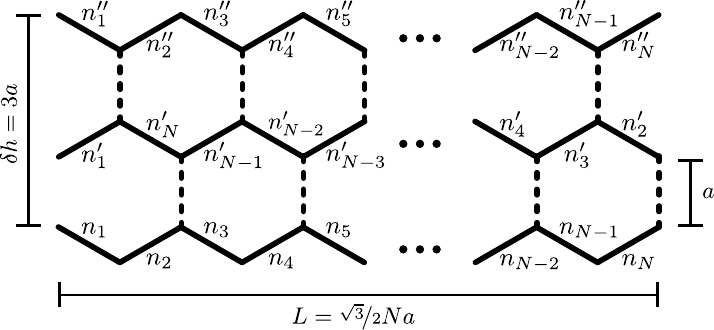}
\caption{
\label{fig: TM illustration}
Bond labels used in the transfer-matrix approach. Measurements are given
relative to the edge length $a$.
}
\end{figure}

We begin by considering a cylindrical lattice of circumference $L$ with $N$
bonds and infinite length, as illustrated in Fig.~\ref{fig: TM illustration}.
We label the coloring configurations of each horizontal row of the system by
$n, n', n'',\ldots \in \B$, where we use the basis set $\B = \{n \in (A,B,C)^N
: n_i \neq n_{i+1} \}$, and we apply periodic boundary conditions $n_i =
n_{i+N}$.

The transfer matrix is a matrix with elements $T_{n,n''}$ that vanish
identically if the stacking of $n$ and $n''$ is not an allowed three coloring
configuration, and that take the value of the sum of the contributions to the
Boltzmann factor of all the bonds on $n$ and between $n$ and $n''$ for an
allowed stacking. With this well-known prescription, the partition function of
the system can be obtained as $Z(L) = \lim_{k \to \infty} {\rm Tr} (T^k)$, from
which we can derive $f(L) = -\lim_{k\to\infty} [\ln{\rm Tr} (T^k)] / (LK)$,
where $K = k \times \delta h$ is the length of the cylinder, and $\delta h$ is
defined in Fig.~\ref{fig: TM illustration}. If $T$ has a unique largest
eigenvalue $\Lambda_0^T$, then $f(L) = -\ln \Lambda_0^T/(L \delta h)$.

With the current choice of orientation, it is clear that the transfer matrix
has geometrically inequivalent indices: row $n$ is not related to row $n'$
by a mere vertical translation, which is why we have so far only defined the
transfer matrix $T$ to act between next-nearest neighboring rows $n$ and
$n''$. $T$ is the standard transfer matrix, and it is positive semi-definite
and symmetric. Here, we choose to proceed by decomposing $T$ into the product
of two semi-transfer matrices $\tau$, which then act between
nearest-neighboring rows $n$ and $n'$. We adopt the labeling scheme shown in
Fig.~\ref{fig: TM illustration}, which treats nearest-neighboring rows
differently. This convention implies that $\sum_{n'} \tau_{n\,n'}\tau_{n'\,n''}
= T_{n\,n''}$. The largest eigenvalue $\Lambda_0^T$ of matrix $T$ is then
equal~\footnote{
Note that $\tau$ is no longer a symmetric matrix, and also not necessarily
positive semi-definite. Nonetheless, for the eigenvector $v$ with largest
(positive) eigenvalue $\Lambda_0^T$ of matrix $T$, the vectors $w_\pm =
\sqrt{\Lambda_0^T} v\pm\tau v$ can be constructed, which are eigenvectors of
$\tau$ with eigenvalues $\pm\sqrt{\Lambda_0^T}$. Also, all other eigenvalues of
$\tau$ have to be smaller in magnitude than $\sqrt{\Lambda_0^T}$. Hence, the
largest eigenvalue of $\tau$ must be $\Lambda_0^\tau = \sqrt{\Lambda_0^T}$.}
to the square of the largest
eigenvalue $\Lambda_0^\tau$ of matrix $\tau$, i.e., we find $\Lambda_0^T =
\left(\Lambda_0^\tau\right)^2$. Therefore by setting the edge length $a=1$, we
find the following relation between the largest eigenvalue of $\tau$ and the
finite-size scaling with bond number $N$,
\begin{equation}
-\frac{2\sqrt{3}}{\pi} \frac{\ln \Lambda_0^{\tau}}{N} =
\gamma_0-\frac{c}{N^2}\,,
\label{eq:finite_size_scaling}
\end{equation}
where $\gamma_0$ is the fitting parameter for the free-energy of the
infinite-size system.

For convenience, we write the semi-transfer matrix as
\begin{equation}
\tau_{n\, n'} = \rho(n) \, \omega(n, n') \, ,
\end{equation}
where $\rho(n)$ accounts for all the Boltzmann weights of the bonds on the
horizontal row $n$, and $\omega(n,n')$ accounts for all the weights of the
vertical bonds connecting rows $n$ and $n'$ (where the numbering according
to Fig.~\ref{fig: TM illustration} is used, and $\omega(n, n') = 0$ for
non-matching $n$ and $n'$). Using Eq.~\eqref{eq:compactified}, this can be
written as:
\begin{equation}
  \rho(n) = \prod_{i=1}^N
            \left[ 1 - (-1)^{\delta_{n_{i-1} n_{i+1}}}~x_{n_i} \right] \, ,
\end{equation}
where $(-1)^{\delta_{n_{i-1} n_{i+1}}} = \pm 1$ if the colors on the bonds
$i-1$ and $i+1$ are different (equal), corresponding to an AFM (FM)
contribution of the interaction term across bond $i$. Note that we are not
checking the validity of the color configuration $n$ because it is inherently
chosen from the set of allowed row colorings $\B$. Similarly, we can write:
\begin{eqnarray}
\omega(n,n')  &=& \!\! \prod_{\substack{j=2\\(\text{even})}}^N
    \left[
		\delta_{n_j n_{N-j+1}'} \delta_{n_{j+1} n_{N-j+2}'}
		(1 + x_{\ell_{n_j,n_{j+1}}})
		\right.
\nonumber \\
&& \quad \;\;\;
		 \left. + \delta_{n_j n_{N-j+2}'} \delta_{n_{j+1} n_{N-j+1}'}
  		(1 - x_{\ell_{n_j,n_{j+1}}})
		 \right]
\, ,
\nonumber \\
\end{eqnarray}
where we introduced the index $\ell_{n_j,n_{j+1}} \in \{A,B,C\}$ such that
$n_j \neq \ell_{n_j,n_{j+1}} \neq n_{j+1}$ (recall that $n_j$ and $n_{j+1}$ are
different by definition). The function $\omega(n, n')$ simultaneously checks
that the two row configurations match one another respecting the color
constraints at all vertices (via the product of delta functions), and
contributes to the corresponding Boltzmann factor.


\subsection{Coloring-sector decomposition and non-propagating sector}
\label{sec:decomposition}

Consider two pairs of two horizontal bonds connected via a vertical bond, e.g.,
$(n_{2m},n_{2m+1})$ and $(n'_{N+2-2m},n'_{N+1-2m})$ in
Fig.~\ref{fig: TM illustration}. Due to the hard coloring constraint, the bonds
of these two pairs must be of the same color, which in turn implies that the
number of $A$, $B$ and $C$ bonds in the basis row is conserved by the action of
the transfer matrix. Thus we can decompose the transfer matrix into sectors
classified by the numbers $N_A$, $N_B$, and $N_C$ of the $A$, $B$, and $C$
bonds respectively within the basis states.

\begin{figure}[tb]
\centering
\includegraphics[width=0.47\columnwidth]{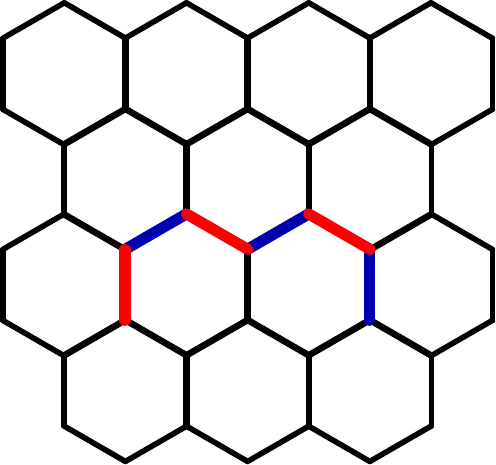}
\hfill
\includegraphics[width=0.47\columnwidth]{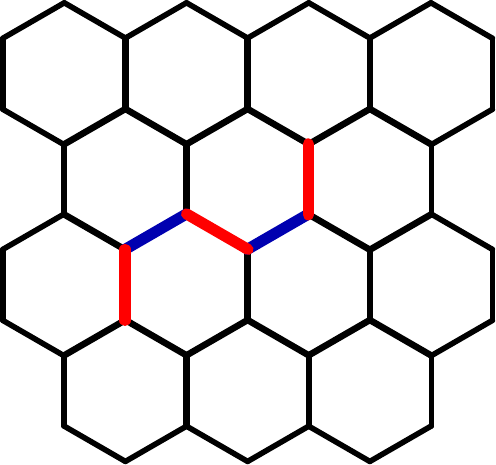}
\caption{
\label{fig:conf_travelling_loop}
(Color online) After crossing between adjacent rows via a vertical bond, a loop
can propagate along the row either an even number of bonds, and cross backward
to the first row via a vertical bond of opposite color (left panel), or it can
propagate an odd number of bonds, and cross forward to the next row via a
vertical bond of the same color (right panel).
}
\end{figure}

This decomposition has important implications, which are clearest in the loop
representation. Observe that each closed $AB$ loop crossing between two
adjacent rows via an $A$ bond must either cross back via a $B$ bond, or cross
forward to the next row via an $A$ bond, as shown in
Fig.~\ref{fig:conf_travelling_loop}.
All closed two-color loops contribute an equal number of bonds of each of the
two colors on a row of vertical bonds. An imbalance in the number of $A$, $B$
and $C$ vertical bonds signals the presence of propagating loops running along
the length of the cylinder. The condition of equal numbers of bonds of each
color on each \textit{vertical} row of bonds is equivalent to the condition of
equal numbers of bonds of each color on each \textit{horizontal} row of bonds,
i.e. $N_A = N_B = N_C$. Hence, all other transfer-matrix sectors that do not
satisfy that condition must contain coloring configurations with propagating
two-color loops.

In this paper, we focus on the transfer-matrix sectors with equal number of
bond colors to study the non-propagating phase of the system. The propagating
phases, where some two-color loops are extended and run either along the
length of the cylinder or wind around it, have their excitations suppressed
in the thermodynamic limit, and hence represent zero-entropy systems that can
only be favored energetically. We therefore expect the phase transition between
the propagating and the non-propagating phases to be of first order. We do not
treat those phases with the finite-size scaling approach, but instead only
compute the line of first-order phase transitions between the propagating and
non-propagating phases in our simulations by comparing the largest eigenvalue
of the non-propagating transfer-matrix sector with the largest eigenvalues of
all other sectors for fixed system size $N=18$.

Finally, for systems with number of bonds on a horizontal row $N$ being not a
multiple of $6$, the condition $N_A = N_B = N_C$ cannot be fulfilled and the
argument presented above forces propagating loops to be present independent of
the values of the coupling constants. These propagating loops appear due to
geometric frustration, which does not play a role in the limit of an
infinite system. Hence, we restrict $N$ for the configurations used in the
transfer-matrix and finite-size calculations to be an integer multiple of $6$.


\subsection{Proof of the irreducibility of the transfer-matrix sectors}
\label{app:proof-irreducible}

\begin{figure*}[p]
  \centering

  \begin{subfigure}{0.4\textwidth}
    \renewcommand*{\thesubfigure}{a}
    \includegraphics[width=\textwidth]
    {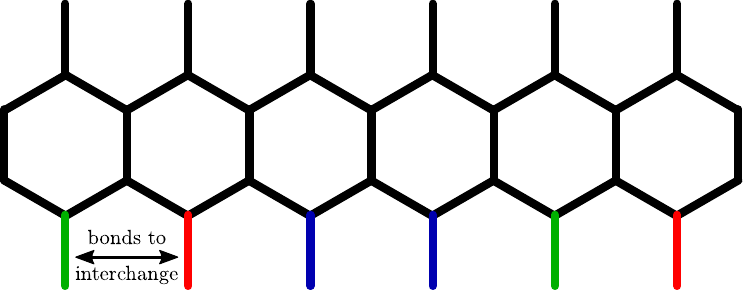}
    \caption{}
  \end{subfigure}~~
  \begin{subfigure}{0.4\textwidth}
    \renewcommand*{\thesubfigure}{a'}
    \includegraphics[width=\textwidth]
    {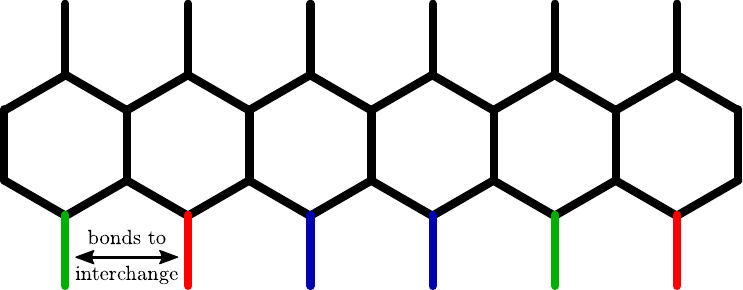}
    \caption{}
  \end{subfigure}

  \begin{subfigure}{0.4\textwidth}
    \renewcommand*{\thesubfigure}{b}
    \includegraphics[width=\textwidth]
    {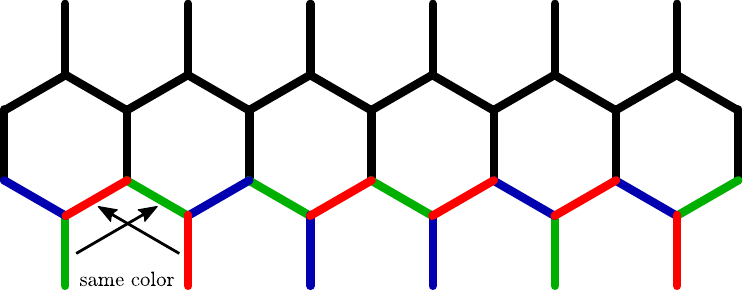}
    \caption{}
  \end{subfigure}~~
  \begin{subfigure}{0.4\textwidth}
    \renewcommand*{\thesubfigure}{b'}
    \includegraphics[width=\textwidth]
    {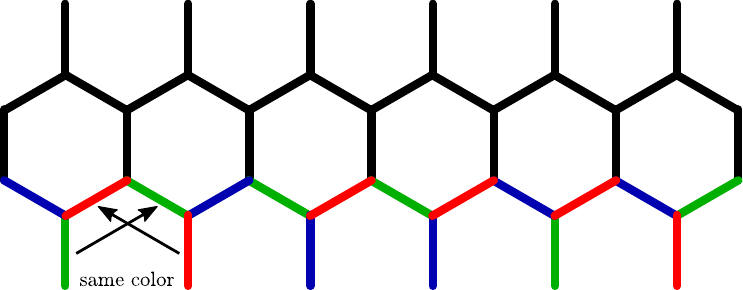}
    \caption{}
  \end{subfigure}

  \begin{subfigure}{0.4\textwidth}
    \renewcommand*{\thesubfigure}{c}
    \includegraphics[width=\textwidth]
    {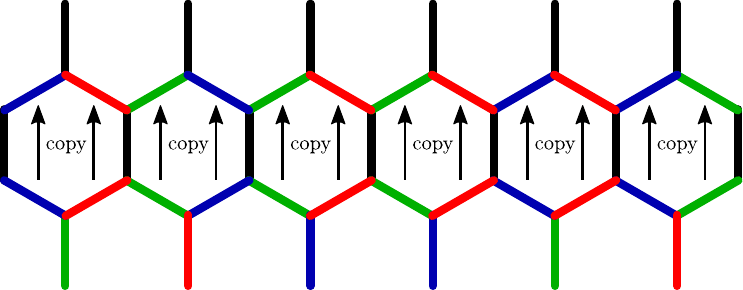}
    \caption{}
  \end{subfigure}~~
  \begin{subfigure}{0.4\textwidth}
    \renewcommand*{\thesubfigure}{c'}
    \includegraphics[width=\textwidth]
    {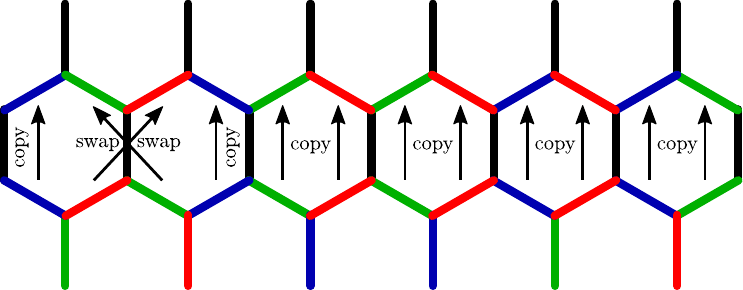}
    \caption{}
  \end{subfigure}

  \begin{subfigure}{0.4\textwidth}
    \renewcommand*{\thesubfigure}{d}
    \includegraphics[width=\textwidth]
    {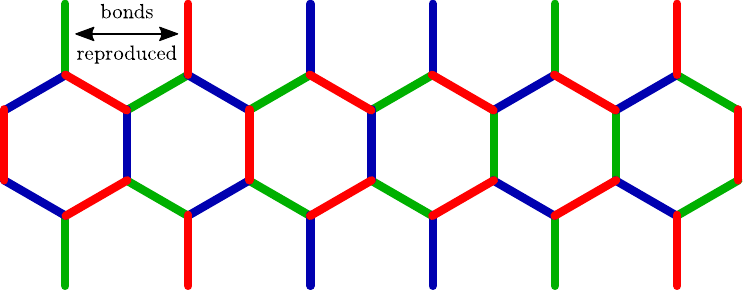}
    \caption{}
  \end{subfigure}~~
  \begin{subfigure}{0.4\textwidth}
    \renewcommand*{\thesubfigure}{d'}
    \includegraphics[width=\textwidth]
    {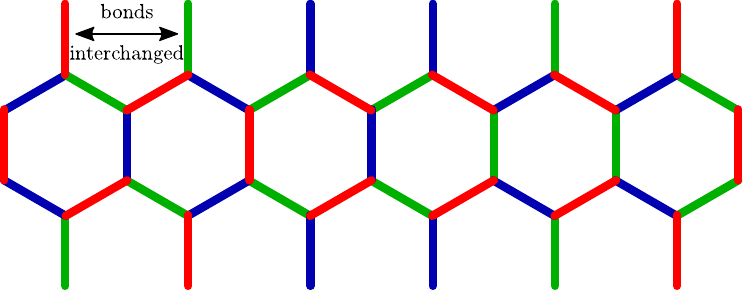}
    \caption{}
  \end{subfigure}

  \vspace{0.5cm}
  \caption{(Color online)
           Two vertical bonds can be interchanged by first reproducing the same
           bonds (left), and then changing the horizontal bonds between those
           two vertical bonds in the process (right).}

  \label{fig:c3-07-swapping-bond-colours}
\end{figure*}

\begin{figure*}[p]
  \centering

  \begin{subfigure}{0.3\textwidth}
    \includegraphics[width=\textwidth]
    {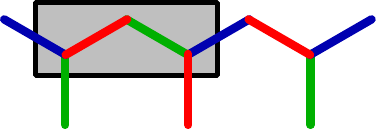}
    \caption{}
  \end{subfigure}~~
  \begin{subfigure}{0.3\textwidth}
    \includegraphics[width=\textwidth]
    {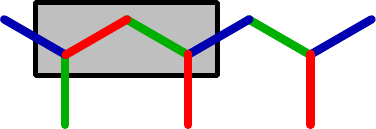}
    \caption{}
  \end{subfigure}

  \begin{subfigure}{0.3\textwidth}
    \includegraphics[width=\textwidth]
    {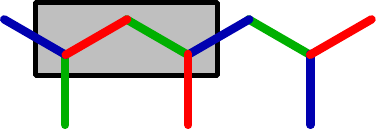}
    \caption{}
  \end{subfigure}~~
  \begin{subfigure}{0.3\textwidth}
    \includegraphics[width=\textwidth]
    {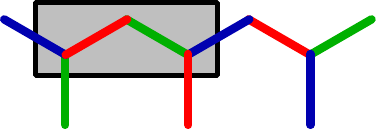}
    \caption{}
  \end{subfigure}

  \vspace{0.5cm}
  \caption  {(Color online)
            To be able to interchange two adjacent vertical bonds as described
            in Fig. \ref{fig:c3-07-swapping-bond-colours}, they have to be
            connected to each other via the same colors on the horizontal row.
            This can always be done if there is at least one vertical bond of
            the other color (blue in this example).}
  \label{fig:c3-08-newbonds}
\end{figure*}

In Sec.~\ref{sec:symmetries}, we will use symmetries of the semi-transfer
matrix to simplify our calculations. This requires that the coloring sectors
defined in Sec.~\ref{sec:decomposition} are irreducible, for which we provide a
proof here.

Irreducibility of each coloring sector is equivalent to the possibility to
connect two given row configurations by a finite number of intermediate steps
of vertical and horizontal rows, which we construct for an arbitrary row
coloring below.
Specifically, we prove that two adjacent vertical bonds can be
interchanged by adding two rows of horizontal and one row of vertical bonds.
Since that operation generates the full permutation group, this is sufficient
as proof.

Start with a configuration of vertical bonds (see
Fig.~\ref{fig:c3-07-swapping-bond-colours} (a)), for which we intend to
interchange two neighboring bonds. We assume those two bonds to be red and
green (without loss of generality). Next, add a set of horizontal bonds (see
Fig.~\ref{fig:c3-07-swapping-bond-colours} (b)) in such a way that the two
vertical bonds that we want to interchange are connected via the same colors,
i.e. again red and green. The fact that this is always possible, is discussed
in the next paragraph. Next, copy all the horizontal bond colorings to the next
horizontal row (see Fig.~\ref{fig:c3-07-swapping-bond-colours} (c)). Finally,
pick all missing bond colors to fulfill the coloring constraint (see
Fig.~\ref{fig:c3-07-swapping-bond-colours} (d)). So far, the original
coloring of the vertical bonds has been reproduced. Next, repeat the four steps
laid out above, but this time, when copying the horizontal bond colorings, swap
the two horizontal bonds connected to the two vertical bonds that we intend to
interchange (see Fig.~\ref{fig:c3-07-swapping-bond-colours} (a') to (d')).
The system ends up with the desired configuration of vertical bonds, where only
two adjacent vertical bonds have been interchanged.

We now have to show that it is always possible to find a horizontal row
configuration that connects the two vertical bonds that we intend to
interchange via the same color, i.e. red and green (see bonds in gray box in
Fig.~\ref{fig:c3-08-newbonds}). This forces the next attached horizontal bond
to be blue. If the next vertical bond is red (green), we are forced to continue
with one green (red) and one blue bond on the horizontal row (see
Fig.~\ref{fig:c3-08-newbonds}(a) and (b)). However, if the next vertical bond
is blue, we can choose the next two horizontal bonds to be either red and green
or green and red (see Fig.~\ref{fig:c3-08-newbonds}(c) and (d)), and
therefore the next bond can be any of the three colors, which puts no further
constraint on all following bonds on the horizontal row. Thus, the statement
holds as long as there exists at least one vertical blue bond.

Therefore, in the case that the vertical bonds have at least one bond of each
color, the sector is irreducible. This is the case when $N_A, N_B, N_C <
\nicefrac{N}{2}$. Otherwise, if for example $N_B = \nicefrac{N}{2}$, the system
is in a state in which all $AC$-loops travel the length of the cylinder. We
assume these sectors to be relevant only for the stripe phase, when it is well
separated (in phase space) from the non-propagating phases that we are focusing
our studies on, and we can therefore neglect the lack of irreducibility for
these sectors.


\subsection{Transfer-matrix symmetries}
\label{sec:symmetries}

The symmetry group of the lattice is generated by two discrete transformations:
rotations $\RR_\pm:n_i \to n_{i\pm 2}$, and inversions $\II_\pm:n_i \to
n_{N+2-i\mp 1}$. The plus (minus) signs have to be taken when acting on
the right (left) increasing states, i.e. when acting on $n$ ($n'$) in
Fig.~\ref{fig: TM illustration}, so that $\tau$ is symmetric with respect to
the symmetry transformations,
\begin{equation}
\label{eq:symmetry_operations}
\tau_{n,n'} = \tau_{\RR_+n,\RR_-n'} \quad \text{and} \quad
\tau_{n,n'} = \tau_{\II_+n,\II_-n'} \, .
\end{equation}
$\RR\in\{\RR_+, \RR_-\}$ and $\II\in\{\II_+, \II_-\}$ generate the full group
of symmetry operations $\A$ with $N$ elements,
\begin{align}
\A = &\left\{\mathbf{1}, \RR, \RR^2, \ldots, \RR^{N/2-1},\right.\nonumber\\
     &\;\;\left. \II, \II\RR, \II\RR^2,\ldots, \II \RR^{N/2-1} \right\} \, ,
\end{align}
and it holds that $\RR^{N/2}=\II^2= \mathbf{1}$, and $\RR \II \RR=\II$.

If $\tau$ was invariant under symmetry operations acting equally on its two
indices, the matrix would naturally decouple into momentum and parity
eigenstates. This is not the case here, as for example $\tau$ couples states
of opposite momenta, which is clear from
$\tau_{n,n'} = \tau_{\RR_+n,\RR_-n'}$. However, the zero
momentum eigensectors still decouple, as we show hereafter.

We define $n_{r\nu} = \RR_+^r \II_+^\nu n_0$ for a fixed $n_0 \in\B$ with $r =
0, 1, 2, \ldots, q-1$, and $\nu = 0,1$, where $q$ is half the length of the
shortest repeating sequence of $n_0$. Analogously, we define $n'_{r'\nu'}$, and
therefore write the elements of $\tau$ as $\tau_{r\nu, r'\nu'} =
\tau_{n_{r\nu}, n_{r'\nu'}}$. The rotational symmetry of $\tau$ takes the form
\begin{equation}
\label{eq:trans_symmetry}
\tau_{r\nu,r'\nu'} = \tau_{r\nu,(-r+r+r')\nu'} = \tau_{0\nu,(r+r')\nu'} =
\tau_{\nu, \nu'}(r+r') \, ,
\end{equation}
and the inversion symmetry takes the form
\begin{align}
   &\tau_{\nu+1, \nu'+1}(r+r') &
 &\stackrel{\mathclap{\eqref{eq:trans_symmetry}}}{=} &
  ~&\tau_{0(\nu+1),(r+r')(\nu'+1)}
 \nonumber\\
 \stackrel{\mathclap{\mathrm{def.}}}{=}
  ~&\tau_{\II_+^{\nu+1} n_0, \RR^{r+r'}_+ \II^{\nu'+1}_+ n_0} &
 &= &
  ~&\tau_{\II_+ \II_+^\nu n_0, \RR^{r+r'}_+ \II_+\II^{\nu'}_+ n_0}
 \nonumber\\
 \stackrel{\mathclap{\eqref{eq:symmetry_operations}}}{=}
  ~&\tau_{\II_+^\nu n_0, \II_- \RR^{r+r'}_+ \II_+ \II^{\nu'}_+ n_0} &
 &\stackrel{\mathclap{(*)}}{=} &
  ~&\tau_{\II_+^\nu n_0, \II_+ \RR^{r+r'-1}_+ \II_+ \II^{\nu'}_+ n_0}
 \nonumber\\
 \stackrel{\mathclap{(**)}}{=}
  ~&\tau_{\II_+^\nu n_0, \RR^{-r-r'+1}_+ \II^{\nu'}_+ n_0} &
 &= &
  ~&\tau_{\nu, \nu'}(-r-r'+1),
\label{eq:invers_symmetry}
\end{align}
where we used that $\II_- = \II_+ \RR_+^{-1} \; (*)$, and
$\II \RR \II = \RR^{-1} \; (**)$.
The transformation into moment and parity eigenstates,
\begin{equation}
v_{k\sigma} = \sum_{r\,\nu} e^{ikr} e^{i\sigma\nu} n_{r\nu} \, ,
\end{equation}
then yields the matrix element,
\begin{align}
     \tau_{k\sigma, k'\sigma'}
  &=~\sum_{r\,r'} \sum_{\nu\,\nu'} e^{-i(kr-k'r')} e^{-i(\sigma\nu
     -\sigma'\nu')} \tau_{r\nu,r'\nu'} \\
  &\stackrel{\mathclap{\eqref{eq:trans_symmetry}}}{=}~
     \delta_{k, -k'} \sum_{r''} e^{-ikr''} \sum_{\nu\,\nu'} e^{-i(\sigma
     \nu - \sigma'\nu')} \tau_{\nu \nu'}(r'') \, , \nonumber\\
\intertext{where we mapped $r \mapsto r+r' = r''$. Hence, the matrix only
couples states of opposite momenta to each other. In particular, the
zero-momentum sector decouples from all other sectors. For $k=k'=0$, it further
holds that}
     \tau_{k\sigma, k'\sigma'}
  &= \sum_{r''} \sum_{\nu\,\nu'} e^{-i(\sigma\nu-\sigma'\nu')}
     \tau_{\nu \nu'}(r'') \\
  &= \sum_{r''} \sum_\nu e^{-i\nu(\sigma-\sigma')} \left[
     \tau_{\nu\nu}(r'') + e^{i\sigma'} \tau_{\nu(\nu+1)}(r'')
     \right] \, , \nonumber
\end{align}
and by using the inversion symmetry of $\tau$ (see
Eq.~\eqref{eq:invers_symmetry}) and letting $r'' \to -r''+1$ in the term with
$\nu = 1$,
\begin{align}
     \tau_{k\sigma, k'\sigma'}
  &= 2\sum_{r''} \sum_\nu e^{-i\nu(\sigma-\sigma')} \left[
     \tau_{00}(r'') + e^{i\sigma'} \tau_{01}(r'')
     \right] \nonumber\\
  &= \delta_{\sigma\sigma'} 2\sum_{r''} \left[
     \tau_{00}(r'') + e^{i\sigma'} \tau_{01}(r'')
     \right] \, .
\end{align}
Thus, also the parity eigensectors decouple.

This result can also be obtained more intuitively by realizing that a symmetric
state $v_0 = v_{k=0,\sigma=0}$ is invariant under symmetry transformations: It
holds that $\RR_- v_0 = \RR_+ v_0$ and $\II_- v_0=\II_+ v_0$. The symmetry
relation for $\tau$ in Eq. \eqref{eq:symmetry_operations} then becomes
\begin{equation}
\tau_{v,v'} = \tau_{\RR_+v,\RR_+v'} \quad \text{and} \quad
\tau_{v,v'} = \tau_{\II_+v,\II_+v'} \, ,
\end{equation}
as long as one of $v$ or $v'$ is in the zero-momentum and zero-parity
eigensector. The result then follows from standard Fourier and
parity-eigenstate transformations.

The fact that the zero-momentum eigensector decouples from all other sectors of
the semi-transfer matrix is important, since it contains the eigenvector
corresponding to the largest eigenvalue. To understand this, let $v$ be an
eigenvector of the semi-transfer matrix $\tau$ with eigenvalue $\lambda$.
A fully-symmetric vector can be defined by
\begin{equation}
  v_\textrm{sym} = \sum_{\Q\in\A} \Q v, \label{eq:gl5}
\end{equation}
which is equally an eigenvector of $\tau$ with eigenvalue $\lambda$. This new
eigenvector could in principle be zero. Since $T$ is non-negative and each
coloring sector is irreducible (see App.~\ref{app:proof-irreducible}), the
Perron-Frobenius theorem insures that all components of the eigenvector
corresponding to the largest eigenvalue are strictly positive. If $v$ is
positive (i.e., all its components are positive), then so must be the
components of the transformed eigenvectors $\Q v$ with $\Q\in\A$. Therefore,
if $\lambda$ is the largest eigenvalue, the symmetric summation in Eq.
\eqref{eq:gl5} has to be non-zero. Thus, for the largest eigenvalue, there
exists at least one eigenvector that is symmetric under all symmetry
transformations $\Q\in\A$ of the semi-transfer matrix $\tau$.

We proceed by transforming $\tau$ into the basis of fully-symmetric vectors.
All other sectors can be neglected, since the symmetric sector contains the
largest eigenvalue. For each subspace that is spanned by the $M$ elements of a
symmetry class $[n] = \{\tilde n : \exists\Q\in\A\text{~so that~}\tilde n=\Q
n\}$, the following matrix of dimension $M \times M$ defines a valid similarity
transformation,
\[
U = \left( \begin{array}{c|c}1 & 1 \,\,\, \cdots \,\,\, 1
    \\\hline
\begin{array}{c} 0 \\ \vdots \\[1.5mm] 0 \end{array} &
\mathlarger{\mathlarger{\mathlarger{\mathbbm{1}}}} \end{array} \right), \,
U^{-1} = \left( \begin{array}{c|c}1 & -1 \,\,\, \cdots \,\,\, -1
    \\\hline
\begin{array}{c} 0 \\ \vdots \\[1.5mm] 0 \end{array} &
\mathlarger{\mathlarger{\mathlarger{\mathbbm{1}}}} \end{array} \right) \, .
\]
Looking at only the first component of this transformed subspace, which
represents the fully-symmetric superposition, it holds that
\begin{align}
  \tau_{[n],[n']} &= [n]~U~\tau~U^{-1}~[n'] \nonumber\\
                  &= \sum_{s\in[n]} \tau_{s,n'} \, .
\end{align}
Thus, to obtain the matrix element between the symmetric superpositions of any
two symmetry classes $[n]$ and $[n']$, we compute the individual matrix element
of each $s$ in the symmetry class $[n]$ with an arbitrary but fixed element
$n'$ of class $[n']$, and sum up all these individual matrix elements. The
spectrum of the matrix remains the same under the similarity transformation.
Therefore, the largest eigenvalue of $\tau$ is the same as the largest
eigenvalue of the matrix in the symmetrized basis.


\subsection{Fitting of finite-size effects}
\label{app:fitting-finite-size}

As explained in the main text, our numerical results are built upon the
finite-size scaling of the reduced free-energy density in
Eq.~\eqref{eq:finite_size_scaling}. This scaling relation holds when mapping
the CFT describing the long-wavelength limit of an infinite system onto a
cylinder. In our simulations, we are only able to access systems of relatively
small size with respect to the lattice spacing. This introduces additional
finite-size effects, which we expect to fade away as $N \to \infty$. In order
to account for this, it is customary to add further terms to the fitting
formula,
\begin{equation}
  -\frac{2\sqrt{3}}{\pi} \frac{\ln \Lambda_0^{\tau}}{N} = \gamma_0 -
  \frac{c}{N^2} + \frac{\gamma_3}{N^3} + \frac{\gamma_4}{N^4} + \ldots \, .
\end{equation}

\begin{figure}[tb]
\includegraphics[width=0.95\columnwidth]{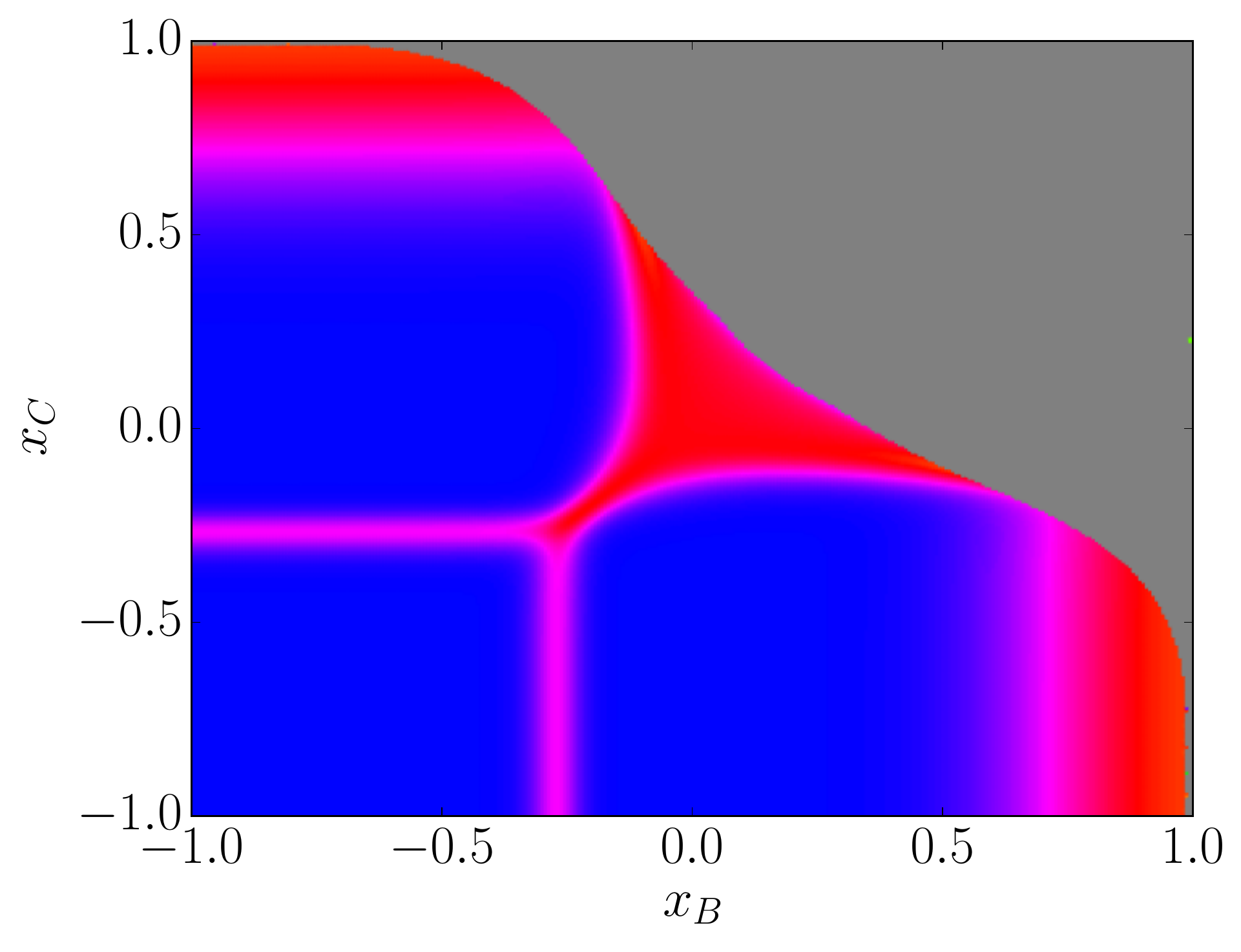}
\\
\includegraphics[width=0.95\columnwidth]{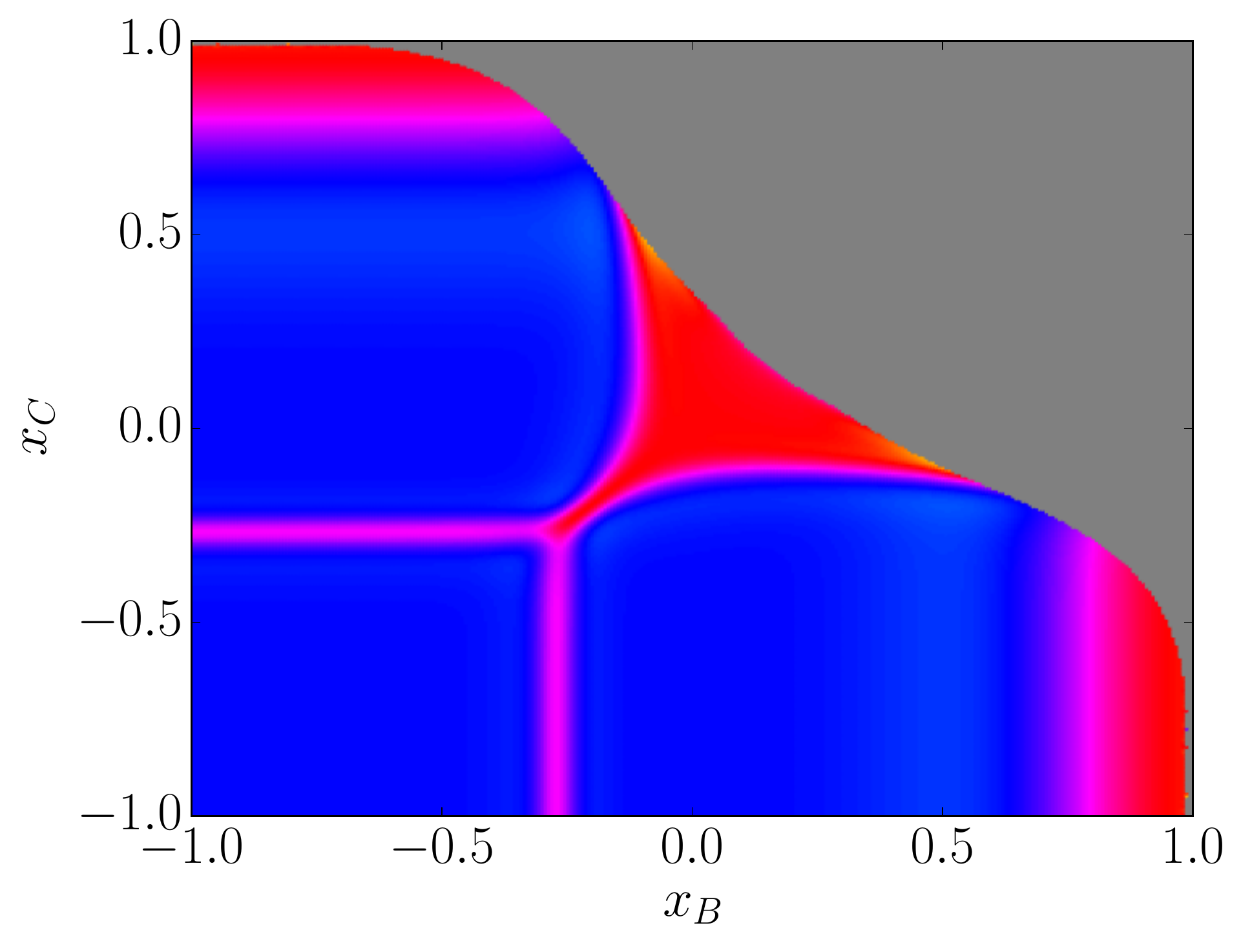}
\caption{\label{fig:comparecnumerics}
(Color online) Top: Results for the central charge for fitting it
with the CFT finite-size scaling using only $c$ and $\gamma_0$ as fitting
parameters. Bottom: The same results using $\gamma_4$ as an additional fitting
parameter. Lattices of size up to $N = 30$ were used to obtain this data. For
the color coding, see Fig.~\ref{fig: JA -inf c Numerics}. Gray regions have
been confirmed to have the largest eigenvalue in one of the propagating
transfer-matrix sectors.}
\end{figure}

We find that our final results depend slightly on what additional fitting
terms we include.
In Fig.~\ref{fig:comparecnumerics}, we present the numerical results
of the central charge for the $x_A \to -1$ plane (compare to
Fig.~\ref{fig: JA -inf c Numerics}) that we obtain by either omitting all
optimization parameters other than $\gamma_0$ and $c$, or by using $\gamma_4$
as an additional fitting parameter, where we set the number of different
system sizes $N$ used equal to the number of optimization parameters.
(Adding $\gamma_3$ or other higher-order fitting terms results in very poor
data with unphysical values of the central charge, and we therefore do not
present it here.)

The additional fitting terms introduce errors, i.e. some slight unphysical
negative values for the central charge. However, they also make transitions
between different values of the central charge sharper.
In the paper we decided to fit the free energy using the optimization
with parameters $\gamma_0$, $c$, and $\gamma_4$.

It is worth mentioning that the data obtained from fits without any additional
optimization parameters ($\gamma_4 = 0$) suggests that the $c=1$ wings
originating from the fully-packed loop model on the $x_B = x_C$ line (see
Sec.~\ref{sec: J_A,J_B = -infinity}) are connected to the $c=1$ region
originating from the fully-frustrated AFM on the triangular lattice in the
corners of the phase diagram (see Sec.~\ref{sec: J_A = -infinity, exact
diagonal}).
This is in contrast to the case when $\gamma_4$ is also used.
For a conclusive answer on this matter, more extensive numerical
calculations are in order, which is beyond the scope of this paper.

Finally, it is important to mention that the numerical results on the
$x_A \to -1$ plane can achieve a higher accuracy than in the rest of
the phase diagram, since all matrix elements with FM $A$ bonds vanish
in this limit. This allows us to significantly reduce the size of the transfer
matrix, and therefore reach larger system sizes with $N$ up to $30$.


\section{SU(3) symmetry in a discrete $\mathcal{S}_3$ system}
\label{app: Read}

At a kagome workshop in 1992, N.~Read presented an argument illustrating how
a continuous SU(3) symmetry can emerge from the inherent discrete
$\mathcal{S}_3$ symmetry in the three coloring model. From this argument, he
further obtained that the model is described in the continuum limit by an
SU(3)$_1$ CFT. We believe that the argument (which was never published) can
help the reader understand the behavior of this model.

The argument begins by noting the equivalence between the phase space of
the three coloring model and the degenerate ground state configurations of
the classical three-state Potts AFM on the kagome lattice at zero temperature.
The classical partition function of the AFM 3-state Potts model is
\begin{subequations}
\begin{eqnarray}
  Z^{(3)}_{\rm Potts}(T) &=&
  \sum_{\{\sigma^{\ }_{i}\}} e^{-\beta H^{(3)}_{\rm Potts}} \, , \\
  H^{(3)}_{\rm Potts} &\equiv&
  J\sum_{\langle ij\rangle} \left( \delta^{\ }_{S_{i}S_{j}} - 1 \right) ,
\end{eqnarray}
where $J>0$, $\langle ij\rangle$ denotes directed nearest-neighbor sites on the
kagome lattice, and the 3-state Potts variables are
\begin{equation}
  S_{i}=1,2,3 \, .
\end{equation}
\end{subequations}
The exchange interaction $J (\delta^{\ }_{S_{i}S_{j}} - 1)$ assumes the value
$-J$ when $S_{i} \neq S_{j}$ and the value $0$ when $S_{i} = S_{j}$. A state
with $S_{i} \neq S_{j}$ for all $\langle ij\rangle$ belongs to the ground state
manifold and has energy $-JN^{\ }_{\langle ij\rangle}$, where $N^{\ }_{\langle
ij\rangle}$ is the number of nearest-neighbor links of the kagome lattice.

\begin{figure}[tb]
  \centering
  \includegraphics[width=0.49\columnwidth]{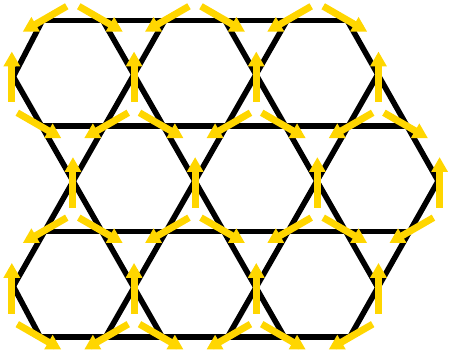}
  \hfill
  \includegraphics[width=0.49\columnwidth]{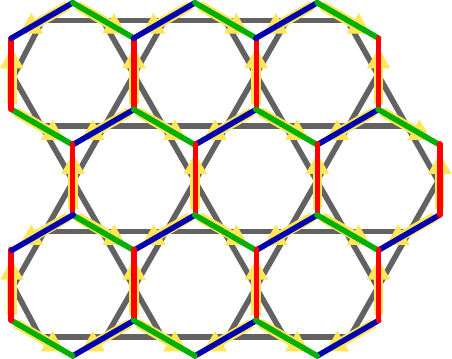}
  \caption{Mapping of a 3-state Potts model ground-state configuration onto
           the three coloring model. Left: An exemplary ground state
           configuration of spins in the 3-state Potts model (also referred to
           as the all-in all-out configuration). Right: The corresponding
           configuration of the three coloring model (observe that this is the
           fully-FM configuration).}
  \label{fig:potts-model}
\end{figure}

At zero temperature,
\begin{eqnarray}
Z^{(3)}_{\rm Potts}(0) &=&
\left[\lim_{\beta\to\infty}e^{\beta JN^{\ }_{\langle ij\rangle}}\right]
\sum_{\{\sigma^{\ }_{i}\}} \prod_{\langle ij\rangle} \left( 1 -
\delta^{\ }_{\sigma^{\ }_{i}\sigma^{\ }_{j}} \right) .
\end{eqnarray}
The ground state manifold of the AFM 3-state Potts model on the kagome lattice
is invariant under the group $\mathcal{S}_3$ of global permutations of the
values $1,2,3$ taken by the Potts spins. One can straightforwardly find a
one-to-one correspondence between this ground state manifold and the phase
space of the three coloring model on the honeycomb lattice by simply drawing
the (dual) hexagonal lattice joining the centers of the triangular plaquettes
in the kagome lattice and identifying the colors $A$, $B$, $C$ with the values
$1$, $2$, $3$ assumed by the Potts variables (see Fig.~\ref{fig:potts-model}).

\begin{figure}[tb]
\centering
\includegraphics[width=0.8\columnwidth]{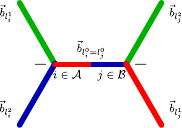}
\caption{
\label{fig: color mismatch}
(Color online) Illustration of the extended phase space of the three coloring
model, which allows color mismatches across the bonds.
}
\end{figure}

The argument to suggest that there exists a hidden SU(3) symmetry in the
3-state Potts AFM at zero temperature is elegantly simple. First of all, let us
extend the phase space allowing for color mismatches across the bonds. Each
vertex has three bonds of different colors departing from it (which we call a
\textit{decorated vertex}), but now bonds connecting two spins can have two
different colors at the two ends (see Fig.~\ref{fig: color mismatch}). The
enlarged phase space is the one obtained by covering the honeycomb lattice with
decorated vertices \textit{independently of one another}, for a total of $6^N$
configurations on a honeycomb lattice of $N$ sites.

One then assigns a three-dimensional (complex) vector
\begin{subequations}
\begin{align}
  \boldsymbol{b}^{\ }_{\ell}\equiv \left( b^{\alpha}_{\ell} \right) =
  \left(
    \begin{array}{c} b^{1}_{\ell} \\ b^{2}_{\ell} \\ b^{3}_{\ell} \end{array}
  \right) \in\mathbb{C}^{3}
\end{align}
to each bond $\ell$ of the lattice,
as illustrated in Figure~\ref{fig: color mismatch},
with the idea of identifying each color with a different component
of the vector: $A$ with the first component, $B$ with the second and
$C$ with the third one.
Using these vectors, one can construct the following terms
\begin{align}
&
\sum^{3}_{\alpha_1=1} \sum^{3}_{\alpha_2=1} \sum^{3}_{\alpha_3=1}
  \varepsilon^{\ }_{\alpha_1,\alpha_2,\alpha_3}
    b^{\alpha_1}_{\ell^{1}_{i}}
      b^{\alpha_2}_{\ell^{2}_{i}}
        b^{\alpha_3}_{\ell^{3}_{i}}
\\
&
\sum^{3}_{\beta_1=1} \sum^{3}_{\beta_2=1} \sum^{3}_{\beta_3 = 1}
  \varepsilon^{\ }_{\beta_1,\beta_2,\beta_3}
    \bar{b}^{\beta_1}_{\ell^{1}_{j}}
      \bar{b}^{\beta_2}_{\ell^{2}_{j}}
        \bar{b}^{\beta_3}_{\ell^{3}_{j}}
\end{align}
associated to vertices $i,j$ belonging to sublattice $\A$ and $\B$
respectively. Here $\alpha_{1,2,3}$ and $\beta_{1,2,3}$ are used to index the
components of vectors $\boldsymbol{b}_{\ell}$ and their complex conjugates
$\bar{\boldsymbol{b}_{\ell}}$. The labels $\ell^{q}_{i}$ and $\ell^{q}_{j}$
refer to the three bonds $q=1,2,3$ departing from vertex $i\in\A$ and $j\in\B$
respectively. This notation, though simple, is clearly redundant, as nearest
neighboring sites share one bond and therefore
\begin{equation}
  \langle i,j \rangle
  \;\Rightarrow\;
  \ell^{q}_{i} \equiv \ell^{q}_{j},
  \qquad
  \exists ! \; q=1,2, \, {\rm or}\, 3 \, ,
\end{equation}
(see Fig.~\ref{fig: color mismatch}).
Note that the Levi-Civita (totally antisymmetric) tensor,
\begin{equation}
\epsilon^{\ }_{\alpha^{\ }_{1}\alpha^{\ }_{2}\alpha^{\ }_{3}} =
  \left\{
    \begin{array}{cl}
      +1 & \textrm{if} \; (\alpha^{\ }_{1},\alpha^{\ }_{2},\alpha^{\ }_{3})
      \textrm{ is an even} \\ & \textrm{permutation of } (1,2,3) \\ & \\
      -1 & \textrm{if} \; (\alpha^{\ }_{1},\alpha^{\ }_{2},\alpha^{\ }_{3})
      \textrm{ is an odd}  \\ & \textrm{permutation of } (1,2,3) \\ & \\
      0  & \hbox{otherwise},
    \end{array}
  \right.
\end{equation}
is used here to ensure that no two colors ($A$, $B$, $C$ $\Leftrightarrow$
$\alpha_1$, $\alpha_2$, $\alpha_3$ $=1,2,3$) are repeated at any vertex. The
product of all the above terms, corresponding to all the sites of the honeycomb
lattice,
\begin{align}
&\prod_{i \in \A}
\left[
  \sum^{3}_{\alpha_1=1} \sum^{3}_{\alpha_2=1} \sum^{3}_{\alpha_3=1}
    \varepsilon^{\ }_{\alpha_1,\alpha_2,\alpha_3}
      b^{\alpha_1}_{\ell^{1}_{i}}
        b^{\alpha_2}_{\ell^{2}_{i}}
          b^{\alpha_3}_{\ell^{3}_{i}}
\right]
\nonumber \\
\times
&\prod_{j \in \B}
\left[
  \sum^{3}_{\beta_1=1} \sum^{3}_{\beta_2=1} \sum^{3}_{\beta_3 = 1}
    \varepsilon^{\ }_{\beta_1,\beta_2,\beta_3}
      \bar{b}^{\beta_1}_{\ell^{1}_{j}}
        \bar{b}^{\beta_2}_{\ell^{2}_{j}}
          \bar{b}^{\beta_3}_{\ell^{3}_{j}}
\right]
\end{align}
gives a sum of terms in one-to-one correspondence with all the $6^N$
configurations of \textit{decorated vertices}.
Each bond $\ell$ contributes a factor given by the product of the
$\boldsymbol{b}^{\ }_{\ell}$ components
\begin{eqnarray}
  b^{\alpha}_{\ell} \bar{b}^{\alpha'}_{\ell} \, ,
  \label{eq: bond b term}
\end{eqnarray}
\end{subequations}
where $\alpha$ and $\alpha'$ are related to the colors of the bond $\ell$
close to its adjacent sites belonging to sublattice $\A$ and to sublattice
$\B$, respectively.

Upon integration over the complex variables
\begin{equation}
\prod_{\alpha,\ell} d b^{\alpha}_{\ell} d \bar{b}^{\alpha}_{\ell} \, ,
\end{equation}
all the terms containing at least one factor~\eqref{eq: bond b term} having
$\alpha \neq \alpha'$ vanish because of complex phase integration. This
eliminates all unwanted configurations where colors mismatch across a given
bond. One can therefore construct the partition function of the original three
coloring model as a sum of integrals over continuous variables. The result
gives the partition function in Read's presentation at the kagome workshop:
\begin{align}
Z^{\ }_{\mathrm{U(3)}} =
& \int \prod_{\ell\in\Lambda} \prod_{\alpha^{\ }_{\ell}=1,2,3}
  \frac{db^{\alpha^{\ }_{\ell}}_{\ell}d\bar{b}^{\alpha^{\ }_{\ell}}_{\ell}}
  {2\pi\mathrm{i}}\exp\left(-\vert b^{\alpha^{\ }_{\ell}}_{\ell} \vert^2\right)
  \nonumber \\ \times
& \prod_{i\in \A}
  \left[
    \sum_{\alpha^{\ }_{1}=1}^{3} \sum_{\alpha^{\ }_{2}=1}^{3}
    \sum_{\alpha^{\ }_{3}=1}^{3} \epsilon^{\ }_{\alpha^{\ }_{1}\alpha^{\ }_{2}
    \alpha^{\ }_{3}} b^{\alpha^{\ }_{1}}_{\ell^{1}_{i}}
    b^{\alpha^{\ }_{2}}_{\ell^{2}_{i}} b^{\alpha^{\ }_{3}}_{\ell^{3}_{i}}
  \right]
\nonumber\\
\times
& \prod_{j\in \B}
  \left[
    \sum_{\beta^{\ }_{1}=1}^{3} \sum_{\beta^{\ }_{2}=1}^{3}
    \sum_{\beta^{\ }_{3}=1}^{3} \epsilon^{\ }_{\beta^{\ }_{1}\beta^{\ }_{2}
    \beta^{\ }_{3}} \bar b^{\beta^{\ }_{1}}_{\ell^{1}_{j}} \bar
    b^{\beta^{\ }_{2}}_{\ell^{2}_{j}} \bar b^{\beta^{\ }_{3}}_{\ell^{3}_{j}}
  \right] ,
\label{eq: b part function}
\end{align}
where $\Lambda$ is the set of all bonds on the honeycomb lattice, and a
convergence Gaussian factor $\exp\left( - \vert b^{\alpha^{\ }_{\ell}}_{\ell}
\vert^2 \right)$ was introduced to normalize the non-vanishing integrals.

In Eq.~\eqref{eq: b part function} we overlooked one important aspect: all
nonvanishing terms in the partition function $Z^{\ }_{\mathrm{U(3)}}$ ought to
be positive, whereas the Levi-Civita tensor elements can take negative values.
The sign of each term in $Z^{\ }_{\mathrm{U(3)}}$ is given by the product of
all elements $\varepsilon^{\ }_{\alpha\beta\gamma}$ appearing in the integrand.
These are nothing but the chirality spins introduced in Sec.~\ref{sec: model},
i.e., the parities of the three colors around each site of the lattice, say
counterclockwise. Whereas this product is not in general positive (see for
instance the recent systematic study in Ref.~\onlinecite{Cepas2017} and
references therein), it is conserved by any loop updates (local or winding).
Indeed, by selecting an alternating coloring path on the lattice, say
$ABABAB...$, and exchanging the two colors along the path $A \leftrightarrow
B$, we flip all the chirality spins along the path. On a lattice with periodic
boundary conditions, the number of sites along the path is \emph{always even}
(for appropriately chosen system sizes that respect sublattice symmetry as well
as color tiling). Therefore, within each sector of phase space identified by
configurations that are connected to one another by loop updates, we have that
$Z^{\ }_{\mathrm{U(3)}} = Z^{(3)}_{\rm Potts}(0)$ or $Z^{\ }_{\mathrm{U(3)}} =
- Z^{(3)}_{\rm Potts}(0)$, and the SU(3) symmetry argument given below holds
separately in each sector.

Moreover, on appropriately commensurate lattices, one can show that the
relevant ordered configurations (staggered, stripe, and columnar) discussed in
the main text satisfy the condition of the product of all chirality spins being
positive, and they are connected to one another via loop updates. According to
Ref.~\onlinecite{Cepas2017}, this sector of phase space is the largest one, and
at the very least N.~Read's construction applies to it.

In order to discuss the symmetries of the partition function
$Z^{\ }_{\mathrm{U(3)}}$ it is convenient to rewrite it as
\begin{align}
  Z^{\ }_{\mathrm{U(3)}} = &
  \int \prod_{\ell\in\Lambda} \frac{d\boldsymbol{b}^{\ }_{\ell} \cdot
    d\bar{\boldsymbol{b}}^{\ }_{\ell}} {2\pi\mathrm{i}}
    \exp\left(-\bar{\boldsymbol{b}}_{\ell} \cdot \boldsymbol{b}_{\ell}\right)
  \nonumber\\ &
  \prod_{i\in A} \left[      \boldsymbol{b}^{\ }_{\ell^{1}_{i}} \cdot \left(
                             \boldsymbol{b}^{\ }_{\ell^{2}_{i}} \wedge
                             \boldsymbol{b}^{\ }_{\ell^{3}_{i}} \right) \right]
  \prod_{j\in B} \left[ \bar{\boldsymbol{b}}^{\ }_{\ell^{1}_{j}}\cdot \left(
                        \bar{\boldsymbol{b}}^{\ }_{\ell^{2}_{j}}\wedge
                        \bar{\boldsymbol{b}}^{\ }_{\ell^{3}_{j}}\right) \right]
  \nonumber\\ = &
  \int \prod_{\ell\in\Lambda} \frac{d\boldsymbol{b}^{\ }_{\ell} \cdot
    d\bar{\boldsymbol{b}}^{\ }_{\ell}} {2\pi\mathrm{i}}
    \exp\left(-\bar{\boldsymbol{b}}_{\ell} \cdot \boldsymbol{b}_{\ell}\right)
  \nonumber\\ &
  \times \exp \left\{
  \sum_{i\in A} \ln \left[   \boldsymbol{b}^{\ }_{\ell^{1}_{i}} \cdot \left(
                             \boldsymbol{b}^{\ }_{\ell^{2}_{i}} \wedge
                             \boldsymbol{b}^{\ }_{\ell^{3}_{i}} \right) \right]
  \right. \nonumber \\ & \qquad\; + \left.
  \sum_{j\in B}\ln\left[\bar{\boldsymbol{b}}^{\ }_{\ell^{1}_{j}}\cdot \left(
                        \bar{\boldsymbol{b}}^{\ }_{\ell^{2}_{j}}\wedge
                        \bar{\boldsymbol{b}}^{\ }_{\ell^{3}_{j}}\right) \right]
  \right\} .
\end{align}
The second equality emphasizes the local and the 3-body nature of the
interaction. The symmetries of $Z^{\ }_{\mathrm{U(3)}}$ are:
\begin{itemize}
\item
  Invariance under local U(1) transformations,
  \begin{align}
    &\!\!\!
    \bar{\boldsymbol{b}}^{\ }_{\ell} \to
    e^{-\mathrm{i}\theta^{\ }_{\ell}} \bar{\boldsymbol{b}}^{\ }_{\ell},
    \qquad
    \boldsymbol{b}^{\ }_{\ell} \to
    e^{+\mathrm{i}\theta^{\ }_{\ell}} \boldsymbol{b}^{\ }_{\ell},
    \\
    &\!\!\!
    \theta^{\ }_{\ell}\in[0,2\pi[ \qquad\quad\, \forall\ell\in\Lambda.
    \nonumber
  \end{align}
  Note that $\boldsymbol{b}^{\ }_{\ell^{1}_{i}} \cdot \left(
  \boldsymbol{b}^{\ }_{\ell^{2}_{i}} \wedge \boldsymbol{b}^{\ }_{\ell^{3}_{i}}
  \right)$ has U(1) charge $+3$ and that
  $\bar{\boldsymbol{b}}^{\ }_{\ell^{1}_{j}} \cdot \left(\bar{\boldsymbol{b}}^
  {\ }_{\ell^{2}_{j}} \wedge \bar{\boldsymbol{b}}^{\ }_{\ell^{3}_{j}}\right)$
  has U(1) charge $-3$.

\item
  Invariance under global SU(3) transformations,
  \begin{align}
    &\!\!\!
    \bar{\boldsymbol{b}}^{\ }_{\ell} \to
    \bar{U} \, \bar{\boldsymbol{b}}^{\ }_{\ell},
    \qquad
    \boldsymbol{b}^{\ }_{\ell} \to U\, \boldsymbol{b}^{\ }_{\ell},
    \\
    &\!\!\!
    \forall\ell\in\Lambda
    \qquad\quad\;\;\;
    U\in\mathrm{SU(3)}.
    \nonumber
  \end{align}
  Note that the measure $\prod_{\ell\in\Lambda} d\boldsymbol{b}^{\ }_{\ell}
  \cdot d\bar{\boldsymbol{b}}^{\ }_{\ell} \, \exp(-\boldsymbol{b}_{\ell} \cdot
  \bar{\boldsymbol{b}}_{\ell})$ is invariant under local U(3) transformations
  \begin{align}
    &\!\!\!
    \bar{\boldsymbol{b}}^{\ }_{\ell} \to
    \bar{U}^{\ }_{\ell} \, \bar{\boldsymbol{b}}^{\ }_{\ell},
    \qquad
    \boldsymbol{b}^{\ }_{\ell} \to
    U^{\ }_{\ell} \, \boldsymbol{b}^{\ }_{\ell},
    \\
    &\!\!\!
    U^{\ }_{\ell}\in\mathrm{U(3)},
    \qquad\;
    \forall\ell\in\Lambda,
    \nonumber
  \end{align}
  whilst the integrand
  \begin{equation}
    \qquad
    \prod_{i\in A}\left[     \boldsymbol{b}^{\ }_{\ell^{1}_{i}} \cdot
                  \left(     \boldsymbol{b}^{\ }_{\ell^{2}_{i}} \wedge
                             \boldsymbol{b}^{\ }_{\ell^{3}_{i}} \right) \right]
    \prod_{j\in B}\left[\bar{\boldsymbol{b}}^{\ }_{\ell^{1}_{j}}\cdot
                  \left(\bar{\boldsymbol{b}}^{\ }_{\ell^{2}_{j}}\wedge
                        \bar{\boldsymbol{b}}^{\ }_{\ell^{3}_{j}}\right) \right]
  \end{equation}
  is invariant under local SU(3) transformations,
  \begin{subequations}
  \begin{align}
    &\!\!\!
    \left(
    \bar{\boldsymbol{b}}^{\ }_{\ell^{1}_{j}},
    \bar{\boldsymbol{b}}^{\ }_{\ell^{2}_{j}},
    \bar{\boldsymbol{b}}^{\ }_{\ell^{3}_{j}}
    \right)
    \to
    \bar{U}^{\ }_{j}
    \left(
    \bar{\boldsymbol{b}}^{\ }_{\ell^{1}_{j}},
    \bar{\boldsymbol{b}}^{\ }_{\ell^{2}_{j}},
    \bar{\boldsymbol{b}}^{\ }_{\ell^{3}_{j}}
    \right),
    \\
    &\!\!\!\;
    U^{\ }_{j}\in\mathrm{SU(3)},
    \qquad
    \forall j\in B,
    \nonumber
    \\
    &\!\!\!
    \left(
    \boldsymbol{b}^{\ }_{\ell^{1}_{i}},
    \boldsymbol{b}^{\ }_{\ell^{2}_{i}},
    \boldsymbol{b}^{\ }_{\ell^{3}_{i}}
    \right)
    \to
    U^{\  }_{i}
    \left(
    \boldsymbol{b}^{\ }_{\ell^{1}_{i}},
    \boldsymbol{b}^{\ }_{\ell^{2}_{i}},
    \boldsymbol{b}^{\ }_{\ell^{3}_{i}}
    \right),
    \\
    &\!\!\!\;
    U^{\ }_{i}\in\mathrm{SU(3)},
    \qquad
    \forall i\in A.
    \nonumber
  \end{align}
  \end{subequations}

\item
  invariance under a transformation that induces the local change
  \begin{equation}
    \qquad\qquad
    \boldsymbol{b}^{\ }_{\ell^{1}_{i}} \cdot \left(
    \boldsymbol{b}^{\ }_{\ell^{2}_{i}} \wedge
    \boldsymbol{b}^{\ }_{\ell^{3}_{i}} \right) \to
  - \boldsymbol{b}^{\ }_{\ell^{1}_{i}} \cdot \left(
    \boldsymbol{b}^{\ }_{\ell^{2}_{i}} \wedge
    \boldsymbol{b}^{\ }_{\ell^{3}_{i}} \right)
  \, .
	\end{equation}
  This can be achieved in many ways. For example, with the following local
	U(3) matrix
  \begin{subequations}
  \begin{equation}
    U^{\ }_{i} =
    \left( \begin{array}{ccc}
    0 & 1 & 0 \\
    1 & 0 & 0 \\
    0 & 0 & 1
    \end{array} \right)
  \end{equation}
  or with the local U(1) transformation
  \begin{equation}
    \theta^{\ }_{\ell^{1}_{i}} =
    \theta^{\ }_{\ell^{2}_{i}} =
    \theta^{\ }_{\ell^{3}_{i}} =
    \frac{\pi}{3}
  \end{equation}
  or with the combined action of a local $V^{\ }_{i}\in\mathrm{SU(3)}$ and a
  local U(1) transformation,
  \begin{align}
    V^{\ }_{i} =&
    \left(
      \begin{array}{ccc}
      0 & e^{+\mathrm{i}\pi/3} & 0 \\
      e^{+\mathrm{i}\pi/3} & 0 & 0 \\
      0 & 0 & e^{+\mathrm{i}\pi/3}
      \end{array}
    \right)
    \\
    \theta^{\ }_{\ell^{1}_{i}} =&
    \theta^{\ }_{\ell^{2}_{i}} =
    \theta^{\ }_{\ell^{3}_{i}} =
    -\frac{\pi}{3} \, .
    \nonumber
  \end{align}
  \end{subequations}
\end{itemize}

Notice that we can equivalently define the vectors $\boldsymbol{b}^{\ }_{\ell}$
in $\mathbb{R}^3$. In this case, the cancellation of color-mismatching
configuration terms is due to vanishing odd Gaussian integrals over real
variables. The resulting $O(3) = \mathbb{Z}_2 \times SO(3)$ symmetry is indeed
a subgroup of $U(3) = U(1) \times SU(3)$ obtained above.

It is interesting to remark how the use of the Levi-Civita tensor, which plays
a key role in uncovering the hidden symmetry, is a non-trivial choice in
Eq.~(\ref{eq: b part function}). Indeed, all we need there is to forbid the
same color to appear twice at the same vertex. This is naturally achieved by a
tensor corresponding to the absolute value of the totally antisymmetric tensor
$\vert\varepsilon_{\alpha\beta\gamma}\vert$. Using instead the Levi-Civita
tensor results in the introduction of spurious negative signs associated to
some of the vertices, which ought to be dealt with carefully, as in our
discussion above. The Levi-Civita tensor (and not its absolute value) is
however key to the hidden SU(3) symmetry.

From the symmetry of this construction, one sees that the height
representation~\cite{Huse1992} takes values in the weight lattice of SU(3). The
global SU(3) symmetry on the lattice is then promoted to a current algebra
symmetry in the continuum, namely that of the well-known Frenkel-Kac
scalar-field representation of `simply-laced' affine Lie algebras. The two
currents of the Cartan subalgebra are $i \partial h_1$ and $i \partial h_2$
(where $h_1$ and $h_2$ are the two components of the scalar field); whereas the
six currents associated with the roots $\vec{G}$ of the reciprocal lattice are
the vertices $\exp(\pm i \vec{G}\cdot\vec{h})$. The normalization due to the
current Lie algebra relations fixes the value of the level $k=1$ as a
byproduct. Similarly to the better known SU(2) case, one thus obtains that the
three coloring model is described in the continuum limit by an
$\mathrm{SU}(3)_{k=1}$ CFT. (A similar argument was later derived and published
independently by Kondev et al~\cite{Kondev1996-1,*Kondev1996-2}.)

We close by noting that the introduction of generic interactions between the
chirality spins explicitly breaks the SU(3) symmetry -- which is the case e.g.,
for the nearest neighbor interactions in Sec.~\ref{sec: model}. Indeed, in
order to identify the Ising spin at a given site of the honeycomb lattice, we
seek a local combination of the $\bar{b}$'s and  $b$'s that is a $U(1)$ singlet
and that picks up a sign under odd permutations of the three colors around the
site.

For instance, the quantity
\begin{subequations}
\begin{align}
  & \det \left(
  \boldsymbol{b}^{\ }_{\ell^{1}_{i}},
  \boldsymbol{b}^{\ }_{\ell^{2}_{i}},
  \boldsymbol{b}^{\ }_{\ell^{3}_{i}}
  \right) \operatorname{per} \left(
  \bar{\boldsymbol{b}}^{\ }_{\ell^{1}_{i}},
  \bar{\boldsymbol{b}}^{\ }_{\ell^{2}_{i}},
  \bar{\boldsymbol{b}}^{\ }_{\ell^{3}_{i}}
  \right),
  \intertext{for $i\in A$ and}
  & \operatorname{per} \left(
  \boldsymbol{b}^{\ }_{\ell^{1}_{j}},
  \boldsymbol{b}^{\ }_{\ell^{2}_{j}},
  \boldsymbol{b}^{\ }_{\ell^{3}_{j}}
  \right) \det \left(
  \bar{\boldsymbol{b}}^{\ }_{\ell^{1}_{j}},
  \bar{\boldsymbol{b}}^{\ }_{\ell^{2}_{j}},
  \bar{\boldsymbol{b}}^{\ }_{\ell^{3}_{j}}
  \right),
\end{align}
\end{subequations}
for $j\in B$, satisfies both conditions. However, if we recall that a permanent
can be expressed as
\begin{equation}
  \operatorname{per} \left(
  \boldsymbol{A},
  \boldsymbol{B},
  \boldsymbol{C}
  \right) =
  \sum_{\alpha=1}^{3}
  \sum_{\beta=1}^{3}
  \sum_{\gamma=1}^{3}
  |\epsilon^{\ }_{\alpha\beta\gamma}|
  A^{\alpha} B^{\beta} C^{\gamma},
\end{equation}
we immediately recognize that the SU(3) symmetry is lost once we introduce such
term in the partition function.


%

\end{document}